\documentclass[reqno,12pt]{amsart}
\usepackage{fullpage}
\usepackage{amsfonts}
\usepackage{amssymb}
\usepackage{lscape}

\numberwithin{equation}{section}
\def\i{{\rm i}}
\def\X{{\bf X}}
\def\prX{{\bf X}^{(1)}}

\def\D#1{D_{#1}}
\def\G{{\mathcal G}}

\def\Si{{\rm Si}}
\def\Ci{{\rm Ci}}

\def\const{{\rm const.}}
\def\Rnum{\mathbb{R}}
\def\Cnum{\mathbb{C}}
\def\Re{{\rm Re}\,}
\def\Im{{\rm Im}\,}

\def\parderop#1{\partial/\partial{#1}}

\def\trans{\rm{trans.}}
\def\scal{\rm{scal.}}
\def\inver{\rm{inver.}}
\def\phas{\rm{phas.}}

\newtheorem{prop}{Proposition}
\newtheorem{thm}{Theorem}

\newtheorem{lem}{Lemma}
\newtheorem{rem}{Remark}

\def\thmref#1{Theorem~\ref{#1}}
\def\lemref#1{Lemma~\ref{#1}}
\def\Ref#1{Ref.\cite{#1}}
\def\secref#1{Sec.~\ref{#1}}

\def\ie/{i.e.}
\def\eg/{e.g.}

\def\endallowdisplaybreaks{}

\begin{document}
\allowdisplaybreaks[3]

\title{Exact solutions of semilinear radial Schr\"odinger equations by separation of group foliation variables}

\author{
Stephen C. Anco$^1$,
Wei Feng$^{2,1}$,
Thomas Wolf$^1$\\
\\\lowercase{\scshape{
${}^1$
Department of Mathematics and Statistics, Brock University\\
St. Catharines, ON L2S3A1, Canada}} \\
\lowercase{\scshape{
${}^2$
Department of Mathematics, Zhejiang University of Technology\\
Hangzhou 310023, China}}}

\begin{abstract}
Explicit solutions are obtained for a class of semilinear radial
Schr\"odinger equations with power nonlinearities in multi-dimensions.
These solutions include new similarity solutions and other new
group-invariant solutions, as well as new solutions that are not
invariant under any symmetries of this class of equations.  Many of
the solutions have interesting analytical behavior connected with
blow-up and dispersion. Several interesting nonlinearity powers arise
in these solutions, including the case of the critical
(pseudo-conformal) power.  In contrast, standard symmetry reduction
methods lead to nonlinear ordinary differential equations for which few if any explicit solutions
can be derived by standard integration methods.
\end{abstract}

\maketitle

\section{\large Introduction}
\label{intro}

An interesting class of nonlinear wave equations
consists of the semilinear Schr\"odinger equations
\begin{equation}\label{nls}
\i u_t=u_{rr} +m u_r/r +k|u|^p u,
\quad
p\neq 0,
\quad
k\neq 0
\end{equation}
for $u(t,r) \in \Cnum$,
where $p\in \Rnum$ is a nonlinearity power, 
$k\in \Rnum$ is the nonlinearity coefficient,
and $m\in \Rnum$ is a spatial-derivative coefficient. 
When $m$ is a positive integer,
this wave equation \eqref{nls} physically describes a general model
for the slow modulation of radial waves
in a weakly nonlinear, dispersive, isotropic medium \cite{Sul}
in $m+1$ dimensions, with radial coordinate $r$.
When $m$ is zero, the equation similarly is a model for
slow modulation of waves in a one-dimensional, weakly nonlinear, dispersive medium,
where $r$ is the full-line coordinate.
In all other cases equation \eqref{nls} can be interpreted instead as
modelling the slow modulation of two-dimensional radial waves in a
planar, weakly nonlinear, dispersive medium containing a point-source
disturbance at the origin, represented by an extra modulation term
$(m-1) u_r/r$ \cite{AncFen}.  
This interpretation can be applied more generally for $m\neq 2$. 
Hereafter we will call \eqref{nls} the radial gNLS (generalized nonlinear Schr\"odinger) equation 
and write
\begin{equation}\label{nls-dim}
m=n-1
\end{equation}
without any restriction on $n\in \Rnum$.

Exact solutions have an important role in the study of 
the radial gNLS equation \eqref{nls-dim}, 
particularly for understanding blow-up, dispersive behaviour, attractors,
and critical dynamics,
as well as for testing numerical solution methods.
Stability and global behaviour of solutions to the initial-value problem 
depend on \cite{Cav,Sul}
the effective dimension $n=m+1$, the nonlinearity power $p$,
and the sign of the interaction coefficient $k$.
Specifically, for $p\geq 4/n>0$ and $k>0$,
some solutions exhibit a finite time blow-up such that
$|u(t,r)|\rightarrow \infty$ as $t\rightarrow T<\infty$.
In the case $p=4/n>0$, which is known as the critical power,
a special class of blow-up solutions is rigorously known to have
the form \cite{Sul}
\begin{equation}\label{crit-blowup}
u(t,r) = (T-t)^{-n/2} U(\xi) \exp(\i(\omega+r^2/4)/(T-t)),
\quad
\xi = r/(T-t),
\quad
\omega\neq 0,
\end{equation}
which is invariant under a certain pseudo-conformal subgroup in
the full symmetry group of equation \eqref{nls},
where $U(\xi)$ satisfies a complex nonlinear second-order ordinary differential equation (ODE)
\begin{equation}\label{crit-blowup-ode}
U'' + (n-1)\xi^{-1} U' +\omega U +k|U|^{4/n}U =0.
\end{equation}
In the supercritical case $p>4/n>0$,
a general class of blow-up solutions is believed \cite{Sul}
to asymptotically approach an exact similarity form
\begin{equation}\label{supercrit-blowup}
u(t,r) = (T-t)^{-1/p} U(\xi) \exp(\i\omega\ln((T-t)/T)),
\quad
\xi = r/\sqrt{T-t},
\quad
\omega\neq 0,
\end{equation}
which is invariant under a certain scaling subgroup in
the full symmetry group of equation \eqref{nls},
where $U(\xi)$ satisfies a more complicated complex nonlinear second-order ODE
\begin{equation}\label{supercrit-blowup-ode}
U'' + ((n-1)\xi^{-1} -\tfrac{1}{2}\i\xi) U' -(\omega +\i/p) U +k|U|^{p}U =0 .
\end{equation}
Both ODEs \eqref{crit-blowup-ode} and \eqref{supercrit-blowup-ode} are, however,
intractable to solve by standard ODE integration techniques \cite{BluAnc,Olv}
such as symmetry reduction and integrating factors.
In fact, as summarized in recent work \cite{AncFen},
the only explicit solutions which are known to-date for $n\neq 1$ ($m\neq 0$)
consist of the obvious constant solution $U=(-\omega/k)^{n/4}\exp(\i\phi)$
for the ODE \eqref{crit-blowup-ode}.

In this paper
we will obtain new explicit exact solutions
to the radial gNLS equation \eqref{nls} for $n\neq 1$ ($m\neq 0$)
by applying a symmetry group method
which has been used successfully in previous work
\cite{AncLiu,AncAliWol1,AncAliWol2}
to find explicit blow-up and dispersive solutions to
semilinear radial wave equations and semilinear radial heat equations
with power nonlinearities in multi-dimensions.
The method uses the group foliation equations
associated with one-dimensional subgroups of the point symmetry group of
a given nonlinear partial differential equation (PDE) \cite{Ovs}.
These equations consist of an equivalent first-order PDE system
whose independent and dependent variables are respectively defined by
the invariants and differential invariants of a given point symmetry subgroup.
Each solution of the system geometrically corresponds to
an explicit one-parameter family of exact solutions of
the original nonlinear PDE, such that the family is closed under
the given point symmetry subgroup, 
which represents a symmetry orbit in the solution space of the PDE.
In the case of a PDE with power nonlinearities,
the form of the resulting group-foliation system
allows explicit solutions to be found by a systematic separation technique
in terms of the group-invariant variables.
We will use an improved version of this technique, which is able to yield
a much wider set of solutions.

Our results include explicit blow-up solutions having the group-invariant forms
\eqref{crit-blowup} and \eqref{supercrit-blowup},
plus explicit blow-up solutions with a non-invariant form,
in addition to explicit dispersive solutions,
explicit standing wave solutions, and explicit monopole solutions.
Among the new solutions,
some are found to hold only for non-integer values of $n-1$ ($=m$),
which we interpret as radial planar solutions in the presence of
a modulation point-source at the origin.

\secref{review} provides a short review of the method of group foliation
and related applications to nonlinear PDEs. 
In \secref{method},
the group-foliation method is applied to the symmetry group of the radial gNLS equation \eqref{nls}.
The improved separation technique used for finding explicit solutions of
the group foliation equations is then introduced in \secref{GHsolns},
and the resulting exact solutions of the radial gNLS equation
along with their basic analytical features
are summarized in \secref{results}.
Finally, some concluding remarks are made in \secref{remarks}.

\section{\large Method of group foliation}
\label{review}

The construction of group foliations using admitted point symmetry groups
for partial differential equations 
is originally due to Lie and Vessiot \cite{Vessiot}
and was revived in its modern form by Ovsiannikov \cite{Ovs}.
An outline of this construction in general goes as follows. 

Let $F=0$ be a given PDE system of order $N\geq 1$ 
with $M\geq 2$ independent variables,
admitting a group $\G$ of point symmetries. 
Then the solution space of $F=0$ is a union of orbits defined by 
the action of $\G$ as a transformation group on solutions. 
Provided that the action of $\G$ is regular and projectable, 
each orbit can be geometrically described as a solution of 
an invariantized system of PDEs, called the group resolving system, 
formulated in terms of the invariants and differential invariants
of the symmetry group $\G$. 
This invariantization of the solution space of $F=0$ 
is most easily carried out in jet space by five main steps:\\
(1) formulate the given PDE system $F=0$ as set of surface equations 
in the jet space of order $N$ using the given variables;\\
(2) express the jet-space variables in all of the surface equations in terms of 
the invariants and differential invariants (up to $N$th order) 
of the symmetry group $\G$;\\
(3) choose $M$ of the lowest order invariantized variables to be 
the new independent variables, 
and take all of the remaining invariantized variables to be 
the new dependent variables;\\
(4) derive the compatibility conditions that come from having the new dependent variables be functions of the new independent variables;\\
(5) append the set of compatibility equations to the set of invariantized surface equations.\\
This set of equations comprises the group-resolving system 
which defines the invariantization of the original PDE system $F=0$. 
Moreover, the original dependent variables can be recovered from the invariantized variables by solving a $\G$-invariant system of differential equations. 
Since the solutions of the group resolving system geometrically correspond to 
the orbits of $\G$ in the solution space of $F=0$, 
each orbit thereby determines a family of solutions to $F=0$ 
such that the family is closed under the action of $\G$. 

The method of group foliation was first applied successfully
to find exact solutions to nonlinear PDEs 
in \Ref{NutkuSheftel1,NutkuSheftel2,SheftelWinternitz,Sheftel1,Sheftel2,Golovin}
when the group $\G$ of point symmetries is infinite-dimensional,
and later it was developed in \Ref{AncLiu,AncAliWol1,AncAliWol2} 
when the point symmetry group $\G$ is finite-dimensional. 

These two basic approaches have been used in many recent papers 
(see, \eg/ \cite{QuZhang,SheftelMakykh}) 
for obtaining exact solutions of nonlinear diffusion equations and nonlinear wave equations. 
In a different direction, 
the formulation of group foliations of nonlinear PDEs 
by using exterior differential systems has been studied 
in \Ref{AndersonFels,Fels}.

\section{\large Symmetries and group foliations}
\label{method}

The group of point symmetries of the radial gNLS equation \eqref{nls}
for $n\neq1$ ($m\neq 0$) is well-known \cite{NikPop,PopKunEsh}
to be generated by the following point transformations acting on $(t,r,u,\bar u)$:
\begin{align}
&\text{ phase rotation } \quad
\X_{\phas} =\i u\parderop{u} -\i \bar{u}\parderop{\bar{u}}
\quad\text{ for all $p$},
\label{phassymm}\\
&\text{ time translation } \quad
\X_{\trans} =\parderop{t}
\quad\text{ for all $p$},
\label{transsymm}\\
&\text{ scaling } \quad
\X_{\scal} =2t\parderop{t} + r\parderop{r} -(2/p) u\parderop{u} -(2/p) \bar u\parderop{\bar{u}}
\quad\text{ for all $p$},
\label{scalsymm}\\
&\text{ inversion } \quad
\begin{aligned}
\X_{\inver} =& t^2\parderop{t} +tr\parderop{r} -(2t/p +\i r^2/4)u \parderop{u}
\\&\qquad
-(2t/p -\i r^2/4)\bar{u}\parderop{\bar{u}}
\quad\text{ only for $p=4/n$}.
\end{aligned}
\label{inversymm}
\end{align}
Note the inversion \eqref{inversymm} is called 
a pseudo-conformal transformation,
and the special power $p=4/n$ for which it exists is commonly called the critical power.

On solutions $u=f(t,r)$ of the radial gNLS equation \eqref{nls},
the one-dimensional symmetry transformation groups 
arising from the separate generators
\eqref{phassymm}--\eqref{inversymm} are given by
\begin{align}
&u=\exp{(\i\phi)}f(t,r),
\label{phas-group}
\\
&u=f(t-\epsilon,r),
\label{trans-group}
\\
&u=\lambda^{-2/p}f(\lambda^{-2}t,\lambda^{-1}r),
\label{scal-group}
\\
&u=(1+\epsilon t)^{-2/p}\exp{(-\i\epsilon r^2/(4+4\epsilon t))} f(t/(1+\epsilon t),r/(1+\epsilon t))
\quad\text{ only for $p=4/n$},
\label{inver-group}
\end{align}
with group parameters
$-\infty<\epsilon<\infty$, $0<\lambda<\infty$, $0 \leq \phi < 2\pi$.
The full transformation group of point symmetries is obtained by 
compositions of these transformations \eqref{phas-group}--\eqref{inver-group}.

A group foliation can be constructed using any linear combination $\X$ of
symmetry generators \eqref{phassymm}--\eqref{inversymm} such that
$\X$ has a regular projectable \cite{Olv} action on $(t,r)$.
In particular, 
\begin{equation}\label{regularX}
\X =c_1\X_{\phas} +c_2\X_{\trans}+c_3\X_{\scal}+c_4\X_{\inver} 
\end{equation}
projects to $(c_2+2c_3 t+c_4 t^2)\parderop{t} + (c_3 r+ c_4tr)\parderop{r}$, 
whose action on $(t,r)$ is regular if and only if 
\begin{equation}\label{regularXcoeffs}
c_2^2+c_3^2+c_4^2 \neq0
\end{equation}
(with $c_4=0$ if $p\neq 4/n$). 
For any symmetry generator $\X$ of the form \eqref{regularX}
with the constraint \eqref{regularXcoeffs}, 
a group foliation consists of converting the radial gNLS equation \eqref{nls}
into a system of first-order equations, 
called the {\em group resolving system},
for the orbits of the one-dimensional symmetry group generated by $\X$.
This system is naturally formulated in terms of a complete set of invariants
$x(t,r)$, $v(t,r,u)$, $\bar{v}(t,r,\bar{u})$
and a complete set of first-order differential invariants
$G(t,r,u,u_t,u_r)$, $\bar{G}(t,r,\bar{u},\bar{u}_t,\bar{u}_r)$,
$H(t,r,u,u_t,u_r)$, $\bar{H}(t,r,\bar{u},\bar{u}_t,\bar{u}_r)$ of $\X$,
which always can be chosen so that the phase-rotation symmetry \eqref{phassymm}
leaves $x$ invariant and acts equivariantly on $v,\bar{v},G,H,\bar{G},\bar{H}$.
As a consequence,
the solution space $\{u=f(t,r)\}$ of the radial gNLS equation \eqref{nls}
can be recovered from the phase-equivariant solution space
$\{(G=g(x,v,\bar{v}),H=h(x,v,\bar{v}))\}$ of the group-resolving system
by integration of the first-order complex differential equations for $u(t,r)$
\begin{equation}\label{GH-odes}
\begin{aligned}
& G(t,r,u,u_t,u_r)=g(x(r,t),v(t,r,u),\bar{v}(t,r,\bar{u})),
\\
& H(t,r,u,u_t,u_r)=h(x(r,t),v(t,r,u),\bar{v}(t,r,\bar{u})),
\end{aligned}
\end{equation}
where this pair of differential equations can be reduced to two quadratures
due to their built-in invariance with respect to 
the two-dimensional symmetry group $\G$ generated by $\X$ and $\X_{\phas}$.
These two quadratures thereby produce a two-parameter family of 
radial gNLS solutions $u=f(t,r,c_1,c_2)$ from each phase-equivariant solution 
$(G=g(x,v,\bar{v}),H=h(x,v,\bar{v}))$ of the group-resolving system. 
Note that the invariance of the differential equations \eqref{GH-odes} 
under phase-rotations is essential for 
having a sufficiently large symmetry group 
to allow integrating them to quadratures. 

We now set up the group-resolving systems for each of 
the symmetry generators given by 
time-translation \eqref{transsymm}, scaling \eqref{scalsymm},
and inversion \eqref{inversymm}.
A general remark is that group-resolving systems arising from
different choices of symmetry groups $\G$ are not related to each other by
a point transformation on $(x,v,\bar{v},G,\bar{G},H,\bar{H})$, 
while the form of any specific group-resolving system 
depends on the complexity of the expressions 
for the symmetry generator $\X$ 
and for the invariants $x,v,\bar{v}$
and differential invariants $G,\bar{G},H,\bar{H}$.
Accordingly, we will leave for other work 
the consideration of group-resolving systems given by
linear combinations of the generators \eqref{phassymm}--\eqref{inversymm},
such as an optimal set with respect to conjugacy in the full symmetry group,
since such systems have a more complicated form that makes it harder to find
explicit solutions by separation of variables. 
(Also see the similar situation for the semilinear wave equation in \Ref{AncLiu}).

\subsection{Time-translation-group resolving system}

To proceed,
we first write down the obvious invariants
\begin{equation}\label{trans-inv}
x = r,
\quad
v = u,
\quad
\bar{v}=\bar{u}
\end{equation}
satisfying $\X_{\trans}x=\X_{\trans}v=\X_{\trans}\bar{v}=0$
and additionally $\X_{\phas}x=0$, $\X_{\phas}v=\i v$, $\X_{\phas}\bar{v}=-\i\bar{v}$.
Similarly, we write down the obvious differential invariants
\begin{equation}\label{trans-diffinv}
G = u_t,
\quad
H = u_r
\end{equation}
satisfying $\prX_{\trans}G=\prX_{\trans}H=0$
and $\prX_{\phas}G=\i G$, $\prX_{\phas}H=\i H$,
where $\prX_{\trans}$ is the first-order prolongation of the time-translation generator \eqref{transsymm}
and $\prX_{\phas}$ is the first-order prolongation of the phase-rotation generator \eqref{phassymm}.
Here $x$, $v$ and $\bar{v}$ are mutually independent,
while $G$ and $H$ are related by equality of mixed $r,t$ derivatives
on $u_{t}$ and $u_{r}$, which gives
\begin{equation}\label{trans-mix}
D_r G = D_t H,
\end{equation}
where $\D{r},\D{t}$ denote total derivatives with respect to $r,t$.
Furthermore,
$v,\bar{v},G,H$ are related through the radial gNLS equation \eqref{nls} by
\begin{equation}\label{trans-nls}
\i G - r^{1-n} D_r(r^{n-1} H)= k v^{1+p/2}\bar v^{p/2}.
\end{equation}
Now we put $G=G(x,v,\bar{v})$, $H=H(x,v,\bar{v})$ into
equations \eqref{trans-mix}--\eqref{trans-nls}
and use equation \eqref{trans-inv} combined with the chain rule
to arrive at a first-order system
\begin{subequations}\label{trans-GH}
\begin{align}
&
G_x + HG_v - GH_v + \bar{H} G_{\bar v} - \bar{G} H_{\bar v}=0
\label{trans-GH1}
\\
&
\i G - (n-1)H/x - H_x - HH_v - \bar{H}H_{\bar v} = kv^{1+p/2}\bar v^{p/2}
\label{trans-GH2}
\end{align}
\end{subequations}
with independent variables $x,v,\bar{v}$,
and dependent variables $G,H$ (and their complex conjugates).
These equations will be called the {\em time-translation-group resolving system}
for the radial gNLS equation \eqref{nls}.

The respective solution spaces of
equation \eqref{nls} and system \eqref{trans-GH}
are related by a group-invariant mapping
that is defined through the invariants \eqref{trans-inv}
and differential invariants \eqref{trans-diffinv},
and that preserves phase-rotation symmetry. 

\begin{lem}\label{trans-map}
Phase-equivariant solutions $(G=g(x,|v|)v,H=h(x,|v|)v)$
of the time-translation-group resolving system \eqref{trans-GH}
are in one-to-one correspondence with
two-parameter families of solutions $u=f(t,r,c_1)\exp(\i c_2)$
of the radial gNLS equation \eqref{nls}
satisfying the time-translation invariance property
\begin{equation}
f(t+\epsilon,r,c_1) = f(t,r,\tilde c_1(\epsilon,c_1)) \exp(\i\tilde c_2(\epsilon,c_2))
\label{trans-orbit}
\end{equation}
(in terms of group parameter $\epsilon$)
for some $\tilde c_1(\epsilon,c_1)$ and $\tilde c_2(\epsilon,c_2)$,
with $\tilde c_1(0,c_1)=c_1$, $\tilde c_2(0,c_2)=0$.
The parameters $c_1,c_2$ arise as the constants of integration of
the pair of first-order DEs
\begin{equation}
u_{r} =h(r,u,\bar{u}), 
\quad
u_{t} =g(r,u,\bar{u})
\label{trans-odes}
\end{equation}
which are invariant under the time-translation symmetry \eqref{transsymm}
and the phase-rotation symmetry \eqref{phassymm}.
\end{lem}

The proof of Lemma~\ref{trans-map} will be given in \secref{trans-GH-solns}.
Through the correspondence stated in this lemma, 
time-translation invariant solutions of the radial gNLS equation \eqref{nls}
with the two-parameter form
\begin{equation}
u =f(r,c_1)\exp(\i c_2)
\label{trans-inv-soln}
\end{equation}
are characterized by the simple condition
\begin{equation}
G=0
\label{trans-inv-GHsoln}
\end{equation}
on phase-equivariant solutions of
the time-translation-group resolving system \eqref{trans-GH}.
This establishes a direct relationship between 
classical symmetry reduction of the radial gNLS equation \eqref{nls} 
under time-translation 
and a reduction of the time-translation-group resolving system \eqref{trans-GH}
under condition \eqref{trans-inv-GHsoln}. 

\begin{lem}
There is a one-to-one correspondence between
two-parameter families of static solutions \eqref{trans-inv-soln}
of the radial gNLS equation \eqref{nls}
and phase-equivariant solutions of the time-translation-group resolving system \eqref{trans-GH}
that satisfy condition \eqref{trans-inv-GHsoln}.
\end{lem}

\subsection{Scaling-group resolving system}

We proceed by writing down the invariants and differential invariants
determined by the scaling generator \eqref{scalsymm}
and its first-order prolongation.
A simple choice of invariants is given by
\begin{equation}\label{scal-inv}
x = t/r^2,
\quad
v = r^{2/p} u,
\quad
\bar{v} = r^{2/p}\bar{u}
\end{equation}
satisfying $\X_{\scal}x=\X_{\scal}v=\X_{\scal}\bar{v}=0$
and $\X_{\phas}x=0$, $\X_{\phas}v=\i v$, $\X_{\phas}\bar{v}=-\i\bar{v}$.
The simplest differential invariants $G(t,r,u_{t})$ and $H(t,r,u_{r})$
satisfying $\prX_{\scal}G=\prX_{\scal}H=0$
and $\prX_{\phas}G=\i G$, $\prX_{\phas}H=\i H$
consist of
\begin{equation}\label{scal-diffinv}
G = r^{2+2/p} u_t,
\quad
H = r^{1+2/p} u_r.
\end{equation}
Here the invariants $x$, $v$ and $\bar{v}$ are again mutually independent,
while the differential invariants $G$ and $H$
are related by equality of mixed $r,t$ derivatives
on $u_{t}$ and $u_{r}$, which gives
\begin{equation}\label{scal-mix}
D_r( r^{-2-2/p}G ) = D_t ( r^{-1-2/p}H ).
\end{equation}
In addition,
$v,\bar{v},G,H$ are related through the radial gNLS equation \eqref{nls} by
\begin{equation}\label{scal-nls}
\i r^{-2-2/p} G - r^{1-n} D_r(r^{n-2-2/p} H) = k r^{-2-2/p} v^{1+p/2}\bar v^{p/2}.
\end{equation}

Now we put $G=G(x,v,\bar{v})$, $H=H(x,v,\bar{v})$ into
equations \eqref{scal-mix}--\eqref{scal-nls}
and apply the chain rule with equation \eqref{scal-inv}
to get a first-order system
\begin{subequations}\label{scal-GH}
\begin{align}
&
2(1+1/p)G + H_x + 2xG_x - (2/p)(vG_v +\bar{v} G_{\bar v})
+ GH_v-  HG_v + \bar G H_{\bar v} - \bar H G_{\bar v}=0
\label{scal-GH1}
\\
&
\i G + (2-n+2/p)H + 2xH_x -(2/p)(vH_v +\bar{v} H_{\bar v})
- HH_v - \bar H H_{\bar v}=kv^{1+p/2}\bar{v}^{p/2}
\label{scal-GH2}
\end{align}
\end{subequations}
with independent variables $x,v,\bar{v}$,
and dependent variables $G,H$ (and their complex conjugates).
These equations will be called the {\em scaling-group resolving system}
for the radial gNLS equation \eqref{nls}.

Similarly to the group foliation based on time-translation,
here the respective solution spaces of
equation \eqref{nls} and system \eqref{scal-GH}
are related by a group-invariant mapping,
as defined through the invariants \eqref{scal-inv} and differential invariants \eqref{scal-diffinv},
preserving phase-rotation symmetry.

\begin{lem}\label{scal-map}
Phase-equivariant solutions $(G=g(x,|v|)v,H=h(x,|v|)v)$
of the scaling-group resolving system \eqref{scal-GH}
are in one-to-one correspondence with
two-parameter families of solutions $u=f(t,r,c_1)\exp(\i c_2)$
of the radial gNLS equation \eqref{nls}
satisfying the scaling invariance property
\begin{equation}
\lambda^{2/p} f(\lambda^2 t,\lambda r,c_1)
= f(t,r,\tilde c_1(\lambda,c_1)) \exp(\i\tilde c_2(\lambda,c_2))
\label{scal-orbit}
\end{equation}
(in terms of group parameter $\lambda$)
for some $\tilde c_1(\lambda,c_1)$ and $\tilde c_2(\lambda,c_2)$,
with $\tilde c_1(1,c_1)=c_1$, $\tilde c_2(1,c_2)=0$,
where $c_1,c_2$ are the constants of integration of
the pair of first-order DEs
\begin{equation}
u_t=r^{-2-2/p}g(t/r^2,r^{2/p}u,r^{2/p}\bar u),
\quad
u_r=r^{-1-2/p}h(t/r^2,r^{2/p}u,r^{2/p}\bar u)
\label{scal-odes}
\end{equation}
which are invariant under the scaling symmetry \eqref{scalsymm}
and the phase-rotation symmetry \eqref{phassymm}.
\end{lem}

The proof of \lemref{scal-map} is given in \secref{scal-GH-solns}.
Through the steps in this proof, 
a simple correspondence can be derived between
similarity solutions of the radial gNLS equation \eqref{nls}
and a particular class of solutions of the scaling-group resolving
system \eqref{scal-GH} as follows.

\begin{lem}\label{similarity-condition}
There is a one-to-one correspondence between
two-parameter families of similarity solutions
\begin{equation}
u =r^{-2/p} f(t/r^2,c_1)\exp(\i c_2)
\label{scal-inv-soln}
\end{equation}
of the radial gNLS equation \eqref{nls}
and phase-equivariant solutions of the scaling-group resolving system \eqref{scal-GH}
that satisfy the condition
\begin{equation}
H+2x G =-(2/p)v .
\label{scal-inv-GHsoln}
\end{equation}
\end{lem}

This correspondence 
establishes a relationship between 
classical similarity reduction of the radial gNLS equation \eqref{nls} 
under scaling symmetry 
and a reduction of the scaling-group resolving system \eqref{trans-GH}
under condition \eqref{scal-inv-GHsoln}.

\subsection{Inversion-group resolving system}

From the inversion generator \eqref{inversymm},
we first write down the mutually independent invariants
\begin{equation}\label{inver-inv}
x = t / r,
\quad
v = r^{n/2} \exp(\i r^2 /(4t)) u,
\quad
\bar{v} = r^{n/2} \exp(-\i r^2 /(4t)) \bar{u}
\end{equation}
satisfying $\X_{\inver}x=\X_{\inver}v=\X_{\inver}\bar{v}=0$
and $\X_{\phas}x=0$, $\X_{\phas}v=\i v$, $\X_{\phas}\bar{v}=-\i\bar{v}$.
Next we write down the simplest choice of mutually independent differential invariants
$G(t,r,u,u_t,u_r)$ and $H(t,r,u,u_t,u_r)$
satisfying $\prX_{\inver}G=\prX_{\inver}H=0$
and $\prX_{\phas}G=\i G$, $\prX_{\phas}H=\i H$:
\begin{subequations}\label{inver-diffinv}
\begin{align}
&
G = r^{2+n/2} \exp(\i r^2/(4t)) \left(u_t + ru_r/t + (n/(2t) + \i r^2/(4t^2))u \right), 
\label{inver-G}\\
&
H =  r^{1+n/2}\exp(\i r^2/(4t)) \left(u_r + \i ru/(2t)\right).
\label{inver-H}
\end{align}
\end{subequations}
These differential invariants are related
by equality of mixed $r,t$ derivatives on $u_{t}$ and $u_{r}$,
and by the radial gNLS equation \eqref{nls},
which yields
\begin{equation}\label{inver-mix}
\begin{aligned}
&
D_r \left( r^{-2-n/2}\exp(-\i r^2/(4t))(G - r^2H/t+(\i r^4/(4t^2)-nr^2/(2t))v) \right)
\\
&
= D_t \left( r^{-1-n/2}\exp(-\i r^2/(4t)) (H-\i r^2 v/(2t)) \right),
\end{aligned}
\end{equation}
and
\begin{equation}\label{inver-nls}
\begin{aligned}
&
\i \big( G -r^2 H/t+(\i r^4/(4t^2)-nr^2/(2t))v \big)
\\
&\qquad
- r^{3-n/2} \exp(\i r^2/(4t))
D_r \left(r^{-2+n/2}(H - \i r^2 v/(2t)) \exp(-\i r^2/(4t)) \right)
\\
&= k v^{1+2/n}\bar v^{2/n}.
\end{aligned}
\end{equation}

Putting $G=G(x,v,\bar{v})$, $H=H(x,v,\bar{v})$ into
equations \eqref{inver-mix}--\eqref{inver-nls}
and applying the chain rule with equation \eqref{inver-inv},
we get a first-order system
\begin{subequations}\label{inver-GH}
\begin{align}
&
(2+n/2)G + xG_x - (n/2)(vG_v +\bar{v} G_{\bar v})
+ GH_v-  HG_v + \bar{G} H_{\bar v } - \bar{H} G_{\bar v}=0
\label{inver-GH1}
\\
&
\i G + (2-n/2)H + xH_x -(n/2)(vH_v + \bar{v} H_{\bar v}) - HH_v - \bar H H_{\bar v}
=kv^{1+2/n}\bar{v}^{2/n}
\label{inver-GH2}
\end{align}
\end{subequations}
with independent variables $x,v,\bar{v}$,
and dependent variables $G,H$ (and their complex conjugates).
These equations will be called the {\em inversion-group resolving system}
for the radial gNLS equation \eqref{nls}.

The respective solution spaces of equation \eqref{nls} and system \eqref{inver-GH}
are related by a group-invariant mapping
that is defined through the invariants \eqref{inver-inv} and differential invariants \eqref{inver-diffinv}
similarly to the group foliations based on time-translation and scaling,
and that preserves phase-rotation symmetry.

\begin{lem}\label{inver-map}
Phase-equivariant solutions $(G=g(x,|v|)v,H=h(x,|v|)v)$
of the inversion-group resolving system \eqref{inver-GH}
are in one-to-one correspondence with
two-parameter families of solutions $u=f(t,r,c_1)\exp(\i c_2)$
of the radial gNLS equation \eqref{nls}
satisfying the pseudo-conformal invariance property
\begin{equation}\label{inver-orbit}
\begin{aligned}
&
(1+\epsilon t)^{-n/2}\exp(-\i \epsilon r^2 /(4+4\epsilon t))
f(t/(1+\epsilon t),r/(1+\epsilon t),c_1)
\\&
= f(t,r,\tilde c_1(\epsilon,c_1)) \exp(\i\tilde c_2(\epsilon,c_2))
\end{aligned}
\end{equation}
(in terms of group parameter $\epsilon$)
for some $\tilde c_1(\epsilon,c_1)$ and $\tilde c_2(\epsilon,c_2)$,
with $\tilde c_1(0,c_1)=c_1$, $\tilde c_2(0,c_2)=0$,
where $c_1,c_2$ are the constants of integration of
the pair of first-order DEs
\begin{subequations}\label{inver-odes}
\begin{align}
& \begin{aligned}&
u_t +(r/t)u_r +(\i r^2/(4t^2) +n/(2t))u
\\&
=r^{-2-n/2}\exp(-\i r^2 /(4t)) g(t/r,r^{n/2}\exp(\i r^2 /(4t))u,r^{n/2}\exp(-\i r^2 /(4t))\bar{u}), 
\end{aligned}
\\
&\begin{aligned}&
u_r +\i r/(2t) u
\\&
=r^{-1-n/2}\exp(-\i r^2 /(4t)) h(t/r,r^{n/2}\exp(\i r^2 /(4t))u,r^{n/2}\exp(-\i r^2 /(4t))\bar{u})
\end{aligned}
\end{align}
\end{subequations}
which are invariant under the inversion (pseudo-conformal) symmetry \eqref{inversymm}
and the phase-rotation symmetry \eqref{phassymm}.
\end{lem}

A proof will be given in \secref{inver-GH-solns}.
Through the steps in the proof, there is a simple correspondence between
pseudo-conformal solutions of the radial gNLS equation \eqref{nls}
and a particular class of solutions of the inversion-group resolving system \eqref{scal-GH}.

\begin{lem}\label{conformal-condition}
There is a one-to-one correspondence between
two-parameter families of pseudo-conformal solutions
\begin{equation}
u =r^{-n/2}\exp(-\i r^2 /(4t)) f(t/r,c_1)\exp(\i c_2)
\label{inver-inv-soln}
\end{equation}
of the radial gNLS equation \eqref{nls}
and phase-equivariant solutions of the inversion-group resolving system \eqref{inver-GH}
that satisfy the condition
\begin{equation}
G=0 . 
\label{inver-inv-GHsoln}
\end{equation}
\end{lem}

This result gives a direct relationship between 
classical reduction of the radial gNLS equation \eqref{nls} 
under the pseudo-conformal symmetry group 
and a reduction of the inversion-group resolving system \eqref{trans-GH}
under condition \eqref{inver-inv-GHsoln}.

\section{\large Solutions of the group-resolving systems}
\label{GHsolns}

We will now explain how a group-invariant map relating
solutions $(G=g(x,v,\bar{v}),H=h(x,v,\bar{v}))$
of a group-resolving system
and two-parameter families of solutions $u=f(t,r,c_1)\exp(\i c_2)$
of the radial gNLS equation \eqref{nls}
arises from integration of the pair of differential equations \eqref{GH-odes}.

Let $y$ be a canonical coordinate given by $\X y=1$ where $\X$ is
the symmetry generator used in constructing the group foliation.
A change of variables in the differential equations \eqref{GH-odes}
via the point transformation $(t,r,u,\bar{u})\rightarrow (y,x,v,\bar{v})$
then yields $v_y=\tilde g(x,v,\bar{v})$ and $v_x=\tilde h(x,v,\bar{v})$,
where $\tilde g$ and $\tilde h$ are each given by
a linear combination of $G$ and $H$ with coefficients depending on $x,v,\bar{v}$.
This pair of first-order differential equations for $v(y,x)$
inherits the invariance of the differential equations \eqref{GH-odes}
with respect to the symmetry generators $\X$ and $\X_{\phas}$,
so consequently, $\tilde g$ and $\tilde h$ can be restricted to have
the phase-equivariant form
$\tilde g(x,v,\bar{v}) = \hat g(x,|v|)v$
and
$\tilde h(x,v,\bar{v}) = \hat h(x,|v|)v$.
Hence the first-order differential equations can be written as a
pair of parametric ODEs
\begin{equation}\label{gh-odes}
v_y=\hat g(x,|v|)v,
\quad
v_x=\hat h(x,|v|)v
\end{equation}
exhibiting explicit symmetry invariance with respect to
$\X=\parderop{y}$ and $\X_{\phas}=\i v\parderop{v} - \i\bar{v}\parderop{\bar v}$.

It is straightforward to integrate these ODEs \eqref{gh-odes}
after $v=A\exp(\i\Phi)$ is expressed in polar form, giving
\begin{align}
&
A_y = A\Re\hat g(x,A),
\quad
\Phi_y = \Im\hat g(x,A), 
\label{APhi-y-ode}\\
&
A_x = A\Re\hat h(x,A),
\quad
\Phi_x = \Im\hat h(x,A).
\label{APhi-x-ode}
\end{align}
In the case when $\Re\hat g \neq0$,
a further change of variables given by the hodograph transformation
$(y,x,A,\Phi)\rightarrow (A,x,y,\Phi)$
converts the polar ODEs \eqref{APhi-y-ode}--\eqref{APhi-x-ode} into the system
\begin{align}
&
y_x = -\Re\hat h(x,A)/\Re\hat g(x,A),
\quad
y_A = 1/(A\Re\hat g(x,A))
\label{y-odes}\\
&
\Phi_x = \Im\hat h(x,A) - \Re\hat h(x,A)\Im\hat g(x,A)/\Re\hat g(x,A),
\quad
\Phi_A = \Im\hat g(x,A)/(A\Re\hat g(x,A))
\label{Phi-odes}
\end{align}
for $y(x,A)$ and $\Phi(x,A)$.
The general solution of this system \eqref{y-odes}--\eqref{Phi-odes} is
given by the line integrals
\begin{align}
&
y = c_1 + \int_\gamma \frac{1}{A\Re\hat g(x,A)} dA
-\frac{\Re\hat h(x,A)}{\Re\hat g(x,A)} dx, 
\label{y-soln}\\
&
\Phi = c_2 + \int_\gamma \frac{\Im\hat g(x,A)}{A\Re\hat g(x,A)} dA
+ \Big( \Im\hat h(x,A) - \frac{\Re\hat h(x,A)\Im\hat g(x,A)}{\Re\hat g(x,A)} \Big) dx
\label{Phi-soln}
\end{align}
in terms of an arbitrary curve $\gamma$ in the $(x,A)$ plane.
These expressions then implicitly determine
\begin{equation}
A=f_1(x,y-c_1),
\quad
\Phi=c_2 + f_2(x,y-c_1), 
\end{equation}
whence
\begin{equation}\label{v-soln-case1}
v=f(x,y-c_1)\exp(\i c_2)
\end{equation}
for some function $f=f_1\exp(\i f_2)$.
Next, in the remaining case $\Re\hat g =0$,
the ODEs \eqref{APhi-y-ode}--\eqref{APhi-x-ode} imply $D_x\hat g=0$
which leads directly to the general solution
\begin{equation}
\Phi = c_2+ \int \Im\hat h(x,A(x)) dx + y \Im\hat g
\label{Phi-soln'}
\end{equation}
with $A(x)$ being determined up to an integration constant $c_1$
from the first-order ODE
\begin{equation}
\frac{dA}{dx} = A\Re\hat h(x,A).
\label{A-soln'}
\end{equation}
Hence
\begin{equation}
A=f_1(x,c_1),
\quad
\Phi=c_2 + f_2(x,c_1) + (\Im\hat g) y, 
\end{equation}
which thereby determines
\begin{equation}\label{v-soln-case2}
v=f(x,y,c_1)\exp(\i c_2),
\quad
|f|_y=0,
\quad
(\arg f)_y=\const
\end{equation}
for some function $f=f_1\exp(\i (f_2+ (\Im\hat g) y))$.
Finally, changing variables $(y,x,v,\bar{v})$ back to $(t,r,u,\bar{u})$
in the formulas \eqref{v-soln-case1} and \eqref{v-soln-case2},
we obtain a two-parameter family of solutions $u=f(t,r,c_1)\exp(\i c_2)$
of the radial gNLS equation \eqref{nls}.

We will next explain the separation technique for finding explicit solutions of
the group-resolving systems \eqref{trans-GH}, \eqref{scal-GH}, \eqref{inver-GH}
for the radial gNLS equation \eqref{nls}.
These systems can be written in the general form
\begin{equation}
\begin{pmatrix}
\Upsilon(G,H) \\ G+\Psi(H)
\end{pmatrix}
=
\begin{pmatrix}
0 \\ -\i kv^{1+p/2}\bar{v}^{p/2}
\end{pmatrix}
\label{GH-sys}
\end{equation}
where $\Psi$ and $\Upsilon$ are quadratically nonlinear first-order
differential operators that possesses the following two properties:
\newline
(1) homogeneity
\begin{subequations}\label{homogen}
\begin{align}
& \Upsilon(\alpha v+\beta v^b\bar{v}^a, \gamma v+\lambda v^b\bar{v}^a)
= \nu v+\mu v^b\bar{v}^a , 
\\
& \Psi(\alpha v+\beta v^b\bar{v}^a)=
\nu v+\mu v^b\bar{v}^a + \epsilon v^{2b-1}\bar{v}^{2a} + \kappa v^{a+b}\bar{v}^{a+b-1} , 
\end{align}
\end{subequations}
with $\alpha$, $\beta$, $\epsilon$, $\kappa$, $\lambda$, $\nu$, $\mu$
denoting functions only of $x$;
\newline
(2) phase invariance
\begin{subequations}\label{phaseinv}
\begin{align}
& \X_{\phas}\Upsilon(v^{a+1}\bar{v}^{a},v^{b+1}\bar{v}^{b})
= \i \Upsilon(v^{a+1}\bar{v}^{a},v^{b+1}\bar{v}^{b}), 
\\
& \X_{\phas}\Psi(v^{b+1}\bar{v}^{b})
=\i \Psi(v^{b+1}\bar{v}^{b}).
\end{align}
\end{subequations}
Based on these properties \eqref{homogen} and \eqref{phaseinv},
a system \eqref{GH-sys} can be expected to have phase-equivariant solutions
given by the separable power form
\begin{align}
&
H=h_1(x) v + h_2(x) v^{a+1}\bar{v}^{a},
\quad
a\neq 0, 
\label{Hansatz}\\
&
G= - \Psi(h_1(x) v + h_2(x) v^{a+1}\bar{v}^{a}) -\i kv^{1+p/2}\bar{v}^{p/2},
\quad
a\neq 0. 
\label{Gansatz}
\end{align}
In particular, the homogeneity properties \eqref{homogen} show that
the $v$ term in $H$ will produce terms in $\Psi(H)$ and $\Upsilon(G,H)$
that contain the same powers $v$,$v^{a+1}\bar{v}^{a}$
already appearing in $H$ and $G$.
Note that these expressions \eqref{Hansatz}--\eqref{Gansatz} for $(H,G)$
have the equivalent phase-equivariant form 
\begin{equation}\label{GHphasequivar}
H=h(x,|v|)v,
\quad
G= g(x,|v|)v
\end{equation}
given by
\begin{equation}\label{hgansatz}
h= h_1 + h_2 |v|^{2a},
\quad
g = -\i k|v|^{p} -\nu-\mu |v|^{2a} -(\epsilon+\kappa)|v|^{4a}, 
\quad
a\neq 0 , 
\end{equation}
where $\nu$, $\mu$, $\epsilon+\kappa$ are certain functions of 
$h_1(x)$ and $h_2(x)$. 

The separation of variables ansatz \eqref{Hansatz}--\eqref{Gansatz} for $(H,G)$
is more general than the two-term ansatzes used in previous work
\cite{AncLiu,AncAliWol1,AncAliWol2}
where the terms in $G$ were restricted to contain the same powers
as the terms in $H$, \eg/
\begin{equation*}
\begin{pmatrix}
H \\ G
\end{pmatrix}
=
\begin{pmatrix}
h_{1} \\ g_{1}
\end{pmatrix} v
+
\begin{pmatrix}
h_{2} \\ g_{2}
\end{pmatrix}
v^{a+1}\bar{v}^{a},
\quad a\neq 0 .
\end{equation*}

Under the improved ansatz \eqref{Hansatz}--\eqref{Gansatz},
a group-resolving system \eqref{GH-sys} will reduce to
a single equation containing the monomial powers
$v$, $v^{a+1}\bar v^{a}$, $v^{2a+1}\bar v^{2a}$, $v^{3a+1}\bar v^{3a}$, $v^{1+p/2}\bar v^{p/2}$, $v^{a+1+p/2}\bar v^{a+p/2}$,
with coefficients that depend on the complex functions $h_1(x),h_2(x)$,
the exponents $a,p$, and the dimension $n$.
From all possible balances among these monomial powers,
five cases arise:
\begin{equation}
a=-p/2;
\quad
a=p/2;
\quad
a=p/4;
\quad
a=p/6;
\quad
a \neq -p/2,p/2,p/4,p/6. 
\end{equation}
In each case, the separate coefficients of the monomials must vanish,
resulting in an overdetermined system of algebraic-differential equations
for the unknowns
\begin{equation}
\Re h_1(x),\Im h_1(x),\Re h_2(x),\Im h_2(x),a,p,n. 
\end{equation}
Such systems can be solved by a systematic integrability analysis,
which we have carried out using the computer algebra program
{\sc Crack} \cite{crack}.
A typical computation is shown in the webpage:
lie.ac.brocku.ca/twolf/papers/AnFeWo2015/readme.txt

\subsection{Results for the time-translation-group resolving system}
\label{trans-GH-solns}

The overdetermined systems of algebraic-differential equations that arise
from reduction of the time-translation-group resolving system \eqref{trans-GH}
under the separation of variables ansatz \eqref{Hansatz}--\eqref{Gansatz}
are found to admit non-zero solutions $(h_1(x),h_2(x))$ only in the cases
$a=p/2$, $a=p/4$, and $a =1/n$.
For $p\neq 0$ and $n\neq1$,
the solutions are given by:
\begin{equation}\label{trans-sol-a}
h_1=h_2=0 ;
\end{equation}
\begin{equation}\label{trans-sol-b}
\begin{aligned}
&
h_1=\Re h_2=0,
\quad
(x^{-1} h_2)'=0,
\\
&
a=1/n,
\quad
n\neq 0 ;
\end{aligned}
\end{equation}
\begin{equation}\label{trans-new-sol-e}
\begin{aligned}
&
h_1 = (2-n)x^{-1},
\quad
\Re h_2=0,
\quad
h_2{}^2 = 2k(2-n)/n,
\\
&
a=p/4,
\quad
p=2/(2-n),
\quad
n\neq 2 ;
\end{aligned}
\end{equation}
\begin{equation}\label{trans-new-sol-f}
\begin{aligned}
&
h_1 = (2-n)x^{-1},
\quad
\Re h_2 =0,
\quad
h_2{}^2 = -k,
\\&
a= p/4,
\quad
p=2(3-n)/(n-2),
\quad
n\neq 2,3 ;
\end{aligned}
\end{equation}
\begin{equation}\label{trans-sol-c}
\begin{aligned}
&
h_1=(2-n)x^{-1},
\quad
\Im h_2=0,
\quad
h_2{}^2=(2-n)k,
\\
&
a= p/4,
\quad
p=2(3-n)/(n-2),
\quad
n\neq 2,3 ;
\end{aligned}
\end{equation}
\begin{equation}\label{trans-sol-d-case1-subcase}
\begin{aligned}
&
h_1=\Im h_2=0,
\quad
h_2'+(n-1)x^{-1} h_2 +k=0,
\\
&
a= -1/2,
\quad
p=-1 ;
\end{aligned}
\end{equation}
\begin{equation}\label{trans-sol-d-case1-generalcase}
\begin{aligned}
&
\Im h_1=\Im h_2=0,
\quad
h_1' +{h_1}^2+(n-1)x^{-1}h_1=0,
\\
&
h_2'+(h_1+(n-1)x^{-1}) h_2 +k=0,
\\
&
a= -1/2,
\quad
p=-1 ;
\end{aligned}
\end{equation}
\begin{equation}\label{trans-sol-d-case2}
\begin{aligned}
&
\Im h_1=\Im h_2=0,
\quad
x^2 h_1'' +(2x^2 h_1 +(n-1)x)h_1'-(n-1)h_1=0,
\\
&
h_2'+ (h_1+(n-1)x^{-1}) h_2 +k=0,
\\
&
a= -1/2,
\quad
p=-1 .
\end{aligned}
\end{equation}

It is simple to integrate the ODEs in equations
\eqref{trans-sol-b}, \eqref{trans-sol-d-case1-subcase},
\eqref{trans-sol-d-case1-generalcase}.
The ODEs in equation \eqref{trans-sol-d-case2}
can be solved in terms of Bessel functions by the following steps.

ODE \eqref{trans-sol-d-case2} for $h_1(x)$ has an integrating factor $x^{-2}$,
which yields
\begin{equation}\label{trans-sol-d-case2-h1-ode}
h_1' +{h_1}^2+(n-1)x^{-1}h_1= C_1
\end{equation}
with $C_1\neq0$.
(Note the case $C_1=0$ is covered by equation \eqref{trans-sol-d-case1-generalcase}.)
This first-order ODE \eqref{trans-sol-d-case2-h1-ode}
is a Riccati equation which can be converted into Bessel's equation
by the transformation $h_1 = (x^{1-n/2} f)'/(x^{1-n/2} f)$, giving
\begin{equation}
x^2 f'' + x f' - (\nu^2 + C_1 x^2) f =0 ,
\quad
\nu=\begin{cases} 1-n/2,  & n\leq 2\\ n/2-1, & n\geq 2\end{cases} . 
\end{equation}
The form of solutions depends on the sign of $C_1$:
\begin{align}
&
f_\nu = C_2 J_{\nu}(\sqrt{-C_1}x) + C_3 Y_{\nu}(\sqrt{-C_1}x)
\text{ for } C_1<0, 
\label{trans-d-riccati-f-bessel}\\
&
f_\nu = C_2 I_{\nu}(\sqrt{C_1}x) + C_3 e^{\nu\pi\i} K_{\nu}(\sqrt{C_1}x)
\text{ for } C_1>0.
\label{trans-d-riccati-f-modified-bessel}
\end{align}
Hence
\begin{equation}\label{trans-d-riccati-h1-f}
h_1 = f_\nu'/f_\nu  \pm \nu x^{-1},
\quad
\nu = \pm(1-n/2) \geq 0
\end{equation}
yields the general solution for $h_1(x)$.
Then ODE \eqref{trans-sol-d-case2} for $h_2(x)$ becomes
\begin{equation}\label{trans-d-riccati-h2-f-ode}
(x^{n/2} f_\nu h_2)' = -k x^{n/2}f_\nu . 
\end{equation}
To integrate this equation \eqref{trans-d-riccati-h2-f-ode},
we consider the cases $C_1>0$ and $C_1<0$ separately.

For the case $C_1<0$, we will use the Bessel function identity
\begin{equation}\label{bessel-id}
z^{\mp\mu}f_{\mu\pm 1}(z) = \mp (z^{\mp\mu}f_\mu(z))' . 
\end{equation}
First apply this identity to the right-hand-side of equation \eqref{trans-d-riccati-h2-f-ode}
with $z=\sqrt{|C_1|}x$ and $\mu=\nu\mp 1=\mp n/2$:
\begin{equation}
x^{n/2}f_\nu = \begin{cases}
x^{1-\nu}f_\nu = -(\sqrt{1/|C_1|} x^{n/2}f_{\nu-1})',
& n\leq 2\\
x^{1+\nu}f_\nu = (\sqrt{|1/C_1|} x^{n/2}f_{\nu+1})',
& n\geq 2
\end{cases} . 
\end{equation}
Then equation \eqref{trans-d-riccati-h2-f-ode} can be directly integrated to get
\begin{equation}\label{trans-d-riccati-h2-f}
h_2 = \begin{cases}
(k/\sqrt{|C_1|}) f_{\nu-1}/f_\nu  + C_4 x^{-n/2}/f_\nu  , & n\leq 2\\
-(k/\sqrt{|C_1|}) f_{\nu+1}/f_\nu  + C_4 x^{-n/2}/f_\nu  , & n\geq 2
\end{cases} . 
\end{equation}
Equation \eqref{trans-d-riccati-h1-f} can be written in a similar form
through the identity \eqref{bessel-id}
with $z=\sqrt{|C_1|}x$ and $\mu=\nu=\pm(1-n/2)$:
\begin{equation}\label{trans-d-riccati-h1-f-ratio}
h_1 =\begin{cases}
\sqrt{|C_1|}f_{\nu -1}/f_\nu, & n\leq 2\\
-\sqrt{|C_1|}f_{\nu +1}/f_\nu, & n\geq 2
\end{cases} . 
\end{equation}
Hence we obtain
\begin{align}
&
h_1 = \pm\sqrt{|C_1|}f_{\mp n/2}/f_{|1-n/2|}, 
\label{trans-d-riccati-h1-soln}\\
&
h_2 = \pm (k/\sqrt{|C_1|}) f_{\mp n/2}/f_{|1-n/2|} + C_4 x^{-n/2}/f_{|1-n/2|}, 
\label{trans-d-riccati-h2-soln}
\end{align}
where the signs are determined by $\pm (1-n/2)\geq 0$,
and where $f_\nu$ is given by the linear combination of
Bessel functions \eqref{trans-d-riccati-f-bessel}.

The case $C_1>0$ is similar but uses the modified Bessel function identity
\begin{equation}\label{modifiedbessel-id}
z^{\mp\mu}f_{\mu\pm 1}(z) = (z^{\mp\mu}f_\mu(z))' .
\end{equation}
This leads to
\begin{align}
&
h_1 = \sqrt{C_1}f_{\mp n/2}/f_{|1-n/2|}, 
\label{trans-d-riccati-h1-soln2}\\
&
h_2 = - (k/\sqrt{C_1}) f_{\mp n/2}/f_{|1-n/2|}+ C_4 x^{-n/2}/f_{|1-n/2|}, 
\label{trans-d-riccati-h2-soln2}
\end{align}
where the signs are again determined by $\pm (1-n/2)\geq 0$,
while $f_\nu$ is given by the linear combination of
modified Bessel functions \eqref{trans-d-riccati-f-modified-bessel}.

Taking into account special cases in the integration of ODEs
\eqref{trans-sol-b}, \eqref{trans-sol-d-case1-subcase},
\eqref{trans-sol-d-case1-generalcase}, \eqref{trans-sol-d-case2},
we obtain 12 solutions for $(h_1(x),h_2(x))$ 
from equations \eqref{trans-sol-a}--\eqref{trans-sol-d-case2}. 
We now list the resulting solutions for $(H,G)$. 

\begin{prop}\label{trans-HG-solns}
For $p\neq0$ and $n\neq 1$,
the ansatz \eqref{Hansatz}--\eqref{Gansatz} yields 
12 phase-equivariant solutions of the time-translation-group resolving system \eqref{trans-GH}:
{\allowdisplaybreaks
\begin{align}
&\label{trans-solGH-a}
\begin{aligned}
H = & 0,
\quad
G = -\i k|v|^{p} v ;
\end{aligned}\\
&\label{trans-solGH-b}
\begin{aligned}
H = & \i C_1 x|v|^{2/n} v,
\quad
G = \left( \i {C_1}^2 x^2 |v|^{4/n} + C_1 n |v|^{2/n} - \i k|v|^{p} \right) v,
\\&
n\neq0,
\quad
C_1\neq0 ;
\end{aligned}\\
&\label{trans-new-solGH-e}
\begin{aligned}
H = & \left( (2-n)x^{-1} \pm \i \sqrt{2k(1-2/n)}|v|^{1/(2-n)} \right)v,
\\
G = & \left( \pm (4-n)\sqrt{2k(1-2/n)}x^{-1}|v|^{1/(2-n)}+\i k(1-4/n)|v|^{2/(2-n)} \right)v,
\\&
p=2/(2-n),
\quad
k(1-2/n)>0,
\quad
n\neq 2 ;
\end{aligned}\\
&\label{trans-new-solGH-f}
\begin{aligned}
H = & \left( (2-n)x^{-1} \pm \i \sqrt{k}|v|^{(3-n)/(n-2)} \right)v,
\quad
G = 0 , 
\\&
p=2(3-n)/(n-2),
\quad
k>0,
\quad
n\neq 2,3 ;
\end{aligned}\\
&\label{trans-solGH-c}
\begin{aligned}
H = & \left( (2-n)x^{-1} \mp \sqrt{(2-n)k} |v|^{(n-3)/(2-n)} \right)v,
\quad
G = 0 , 
\\&
p=2(3-n)/(n-2),
\quad
k(2-n)>0,
\quad
n\neq 2,3 ;
\end{aligned}\\
&\label{trans-solGH-d-case1-subcase-general-sol1}
\begin{aligned}
H = & ( -(k/n)x + C_1 x^{1-n} )|v|^{-1} v,
\quad
G = & 0 , 
\\&
p=-1,
\quad
n\neq0 ;
\end{aligned}\\
&\label{trans-solGH-d-case1-subcase-general-sol2}
\begin{aligned}
H = & x(C_1 - k\ln x)|v|^{-1} v,
\quad
G = 0 , 
\\&
p=-1,
\quad
n=0 ;
\end{aligned}\\
&\label{trans-solGH-d-case1-general-sol1}
\begin{aligned}
H = & \left( (2-n)(x + C_1 x^{n-1})^{-1}(1 + (C_2+ (k/(2n))x^2)|v|^{-1}) -(k/n)x |v|^{-1} \right)v ,
\quad
G = 0 , 
\\&
p=-1,
\quad
n\neq0,2 ;
\end{aligned}\\
&\label{trans-solGH-d-case1-general-sol2}
\begin{aligned}
H = &
\left( x(x^2+C_1)^{-1}(2 -(kC_1\ln x + C_2 )|v|^{-1}) -(k/2)x |v|^{-1} \right)v ,
\quad
G = 0 , 
\\&
p=-1,
\quad
n=0 ;
\end{aligned}\\
&\label{trans-solGH-d-case1-general-sol3}
\begin{aligned}
H = &
\left( (\ln x +C_1)^{-1} x^{-1}(1 +(C_2 + (k/4) x^2)|v|^{-1}) -(k/2)x |v|^{-1} \right)v ,
\quad
G = 0 , 
\\&
p=-1,
\quad
n=2 ;
\end{aligned}\\
& \label{trans-solGH-d-case2-general-1}
\begin{aligned}
H = &
\pm\sqrt{C_1}\left( C_2J_{|1-n/2|}(\sqrt{C_1}x) + C_3 Y_{|1-n/2|}(\sqrt{C_1}x) \right)^{-1}
\times\\&\qquad
\left( ( C_2J_{\mp n/2}(\sqrt{C_1}x) + C_3 Y_{\mp n/2}(\sqrt{C_1}x) )
(1 +(k/C_1) |v|^{-1} ) + C_4 x^{-n/2} |v|^{-1} \right)v, 
\\
G = &
\i C_1v,
\\&
p=-1,
\quad
\pm(1-n/2)\geq 0,
\quad
C_1>0 ;
\end{aligned}\\
& \label{trans-solGH-d-case2-general-2}
\begin{aligned}
H = &
\sqrt{C_1}\left( C_2I_{|1-n/2|}(\sqrt{C_1}x) + C_3e^{\i\pi |1-n/2|} K_{|1-n/2|}(\sqrt{C_1}x) \right)^{-1}
\times\\&\quad
\left( ( C_2I_{\mp n/2}(\sqrt{C_1}x) + C_3 e^{\mp \i \pi n/2}K_{\mp n/2}(\sqrt{C_1}x) )( 1 -(k/C_1) |v|^{-1} ) + C_4 x^{-n/2} |v|^{-1} \right)v, 
\\
G = &
-\i C_1 v,
\\&
p=-1,
\quad
\pm(1-n/2)\geq 0,
\quad
C_1>0 . 
\end{aligned}
\end{align}
\endallowdisplaybreaks}
Solutions \eqref{trans-new-solGH-f}--\eqref{trans-solGH-d-case1-general-sol3}
satisfy the translation-invariance condition \eqref{trans-inv-GHsoln}.
\end{prop}

For each phase-equivariant solution $(G=g(x,|v|)v,H=h(x,|v|)v)$ of
the time-translation-group resolving system \eqref{trans-GH},
the differential invariants \eqref{trans-diffinv} of $\X_{\trans}$
yield a pair of DEs \eqref{trans-odes} which take the form 
\begin{equation}
v_y=u_t=G=g(x,|v|)v,
\quad
v_x=u_r=H=h(x,|v|)v
\end{equation}
expressed in terms of the invariants $x=r$, $v=u$, $\bar{v}=\bar{u}$
and the canonical coordinate $y=t$ of $\X_{\trans}$.
These DEs determine a two-parameter family of solutions
$u=f(t,r,c_1)\exp(\i c_2)$ of the radial gNLS equation \eqref{nls},
corresponding to orbits of the two-dimensional symmetry group $\G$ 
generated by $\X_{\trans}$ and $\X_{\phas}$.
In the notation \eqref{gh-odes}, with $\hat g=g$ and $\hat h=h$,
the explicit polar form $u=A\exp(\i \Phi)$ of the solution families is given by
the line integral formula \eqref{y-soln}--\eqref{Phi-soln}
in the case $\Re\hat g\neq 0$
and the integration formula \eqref{Phi-soln'}--\eqref{A-soln'}
in the case $\Re\hat g=0$.
Hence, these formulas establish a group-invariant mapping from
phase-equivariant solutions of the time-translation-group resolving system \eqref{trans-GH}
into a class of solutions of the radial gNLS equation \eqref{nls} satisfying
the time-translation invariance property \eqref{trans-orbit},
where
\begin{align}
&
\tilde c_1 = c_1-\epsilon,
\quad
\tilde c_2 = 0,
\quad
\text{ when } \Re\hat g\neq 0, 
\label{c1c2-case1}\\
&
\tilde c_1 = c_1,
\quad
\tilde c_2 = (\Im\hat g)\epsilon,
\quad
\text{ when } \Re\hat g= 0.
\label{c1c2-case2}
\end{align}
An inverse mapping can be constructed in each case by the following steps.

Consider the case of solutions of the radial gNLS equation \eqref{nls}
having the two-parameter form $u=f(t+c_1,r)\exp(\i c_2)$
with $|f|_{c_1}\neq 0$ holding in some open domain in the $(t,r)$ plane.
Under the change of variables $t=y$, $r=x$, $u=v$,
each such solution determines a function \eqref{v-soln-case1}
from which the differential invariants \eqref{trans-diffinv} of $\X_{\trans}$
are given by
\begin{equation}\label{trans-vy-vx}
G=u_t=v_y=g(y+c_1,x)v,
\quad
H=u_r=v_x=h(y+c_1,x)v
\end{equation}
as written in terms of the functions
$g= (\ln|v|+\i\arg v)_y$ and $h= (\ln|v|+\i\arg v)_x$.
These two functions are related by
the differential identity
\begin{equation}\label{trans-vy-vx-id}
D_x G= D_y H
\end{equation}
and the radial gNLS equation
\begin{equation}\label{trans-vy-vx-nls}
\i G = x^{1-n} D_x(x^{n-1} H) + k v^{1+p/2}\bar v^{p/2} . 
\end{equation}
Now, from the relation $|v|=|f(y+c_1,x)|$,
since $|f|_{c_1}=|f|_y\neq 0$ holds locally in the $(y,x)$ plane,
the implicit function theorem can be used to express
$y+c_1= F(x,|v|)$ in terms of some function $F$.
When this expression is substituted into $g$ and $h$, 
they each become a function of just $x$ and $|v|$.
Hence the differential invariants \eqref{trans-vy-vx}
become phase-equivariant functions of $x,v,\bar{v}$,
which satisfy the time-translation-group resolving system \eqref{trans-GH}
as consequence of equations \eqref{trans-vy-vx-id}--\eqref{trans-vy-vx-nls}.

Finally, consider the case of solutions of the radial gNLS equation \eqref{nls}
given by the two-parameter form $u=f(t,r,c_1)\exp(\i c_2)$
with $|f|_t=0$, $(\arg f)_t=\const$,
and $|f|_{c_1}\neq 0$ holding in some open domain in the $(t,r)$ plane.
Each such solution determines a function \eqref{v-soln-case2}
after the change of variables $t=y$, $r=x$, $u=v$.
The differential invariants \eqref{trans-diffinv} of $\X_{\trans}$
again have the form \eqref{trans-vy-vx} in terms of two functions
$g= (\ln|v|+\i\arg v)_y$ and $h= (\ln|v|+\i\arg v)_x$,
satisfying equations \eqref{trans-vy-vx-id} and \eqref{trans-vy-vx-nls}.
Now, since $|f|_t=|f|_y=0$ and $|f|_{c_1}\neq 0$
hold locally in the $(y,x)$ plane,
the implicit function theorem can be applied to 
the relation $|v|=|f(y,x,c_1)|$,
giving $c_1= F(x,|v|)$ in terms of some function $F$.
This expression allows $c_1$ to be eliminated from $h$
which then yields a function of just $x$ and $|v|$,
while $g=(\ln|f|+\i\arg f)_y$ reduces to a constant 
as a consequence of $|f|_y=0$ and $(\arg f)_y=\const$.
The differential invariants \eqref{trans-vy-vx} thereby
become phase-equivariant functions of $x,v,\bar{v}$,
which satisfy the time-translation-group resolving system \eqref{trans-GH}
due to equations \eqref{trans-vy-vx-id}--\eqref{trans-vy-vx-nls}.

This completes the proof of \lemref{trans-map}.

\subsection{Results for the scaling-group resolving system}
\label{scal-GH-solns}

The overdetermined systems of algebraic-differential equations obtained by
reduction of the scaling-group resolving system \eqref{scal-GH}
under the separation of variables ansatz \eqref{Hansatz}--\eqref{Gansatz}
are found to admit non-zero solutions $(h_1(x),h_2(x))$
only in the cases $a=p/2$, $a=p/4$, $a\neq -p/2,p/2,p/4,p/6$.
For $p\neq 0$ and $n\neq1$,
this yields the solutions:
\begin{equation}\label{scal-sol-a}
h_1=h_2=0 ;
\end{equation}
\begin{equation}\label{scal-sol-b}
h_1=-\i/(2x),
\quad
h_2=0 ;
\end{equation}
\begin{equation}\label{scal-sol-c}
\begin{aligned}
&
h_1=\Re h_2 =0,
\quad
h_2'=0,
\\
&
a=1/n,
\quad
p=2/n,
\quad
n\neq 0 ;
\end{aligned}
\end{equation}
\begin{equation}\label{scal-sol-h}
\begin{aligned}
&
h_1=2-n,
\quad
\Re h_2=0,
\quad
h_2{}^2=2k(2-n)/n,
\\
&
a=p/4,
\quad
p=2/(2-n),
\quad
n\neq 2 ;
\end{aligned}
\end{equation}
\begin{equation}\label{scal-sol-f}
\begin{aligned}
&
h_1=2-n,
\quad
\Im h_2=0,
\quad
h_2{}^2=k(2-n),
\\
&
a=p/4,
\quad
p=2(3-n)/(n-2),
\quad
n\neq 2,3 ;
\end{aligned}
\end{equation}
\begin{equation}\label{scal-sol-i}
\begin{aligned}
&
h_1=2-n,
\quad
\Re h_2=0,
\quad
h_2{}^2=-k,
\\
&
a=p/4,
\quad
p=2(3-n)/(n-2),
\quad
n\neq 2,3 ;
\end{aligned}
\end{equation}
\begin{equation}\label{scal-sol-d}
\begin{aligned}
&
h_1=2-n,
\quad
h_2=-k/2,
\\
&
a=-1/2,
\quad
p=-1 ;
\end{aligned}
\end{equation}
\begin{equation}\label{scal-sol-e}
\begin{aligned}
&
h_1=0,
\quad
h_2=-k/n,
\\
&
a=-1/2,
\quad
p=-1,
\quad
n\neq 0,2 ;
\end{aligned}
\end{equation}
\begin{equation}\label{scal-sol-k}
\begin{aligned}
&
h_1=-1,
\quad
h_2=-\i k x,
\\
&
a=-1/2,
\quad
p=-1,
\quad
n=3 ;
\end{aligned}
\end{equation}
\begin{equation}\label{scal-sol-l}
\begin{aligned}
&
h_1=-\i/(2x),
\quad
h_2=k/4,
\\
&
a=-1/2,
\quad
p=-1,
\quad
n=-4 ;
\end{aligned}
\end{equation}
\begin{equation}
\begin{aligned}\label{scal-sol-s}
&
h_1=6 - \i/(2x),
\quad
h_2 = -k/2,
\\
&
a=-1/2,
\quad
p=-1,
\quad
n=-4 ;
\end{aligned}
\end{equation}
\begin{equation}
\begin{aligned}\label{scal-sol-m}
&
h_1=2-n - \i/(2x),
\quad
\Im h_2=0,
\quad
h_2{}^2 = -kn(n+2),
\\
&
a=1/n,
\quad
p=4/n,
\quad
n^2-n-4=0 ;
\end{aligned}
\end{equation}
\begin{equation}
\begin{aligned}\label{scal-solh-o}
&
h_1 = 2-n - \i/(2x),
\quad
\Re h_2 = 0,
\quad
h_2{}^2 = -k,
\\
&
a=1/n,
\quad
p = 4/n,
\quad
n^2-n-4=0,
\quad
k>0 ;
\end{aligned}
\end{equation}
\begin{equation}\label{scal-sol-n}
\begin{aligned}
&
h_1 = 2/3 - \i/(2x),
\quad
\Re h_2 = 0,
\quad
h_2{}^2 = k,
\\
&
a=3/4,
\quad
p = 3,
\quad
n=4/3,
\quad
k<0 ;
\end{aligned}
\end{equation}
\begin{equation}\label{scal-solh-p}
\begin{aligned}
&
h_1 = -1 - \i/(2x),
\quad
h_2 = -\i (2k/5)x,
\\
&
a=-1/2,
\quad
p=-1,
\quad
n=3 ;
\end{aligned}
\end{equation}
\begin{equation}\label{scal-solh-r}
\begin{aligned}
&
\Im h_1 = -1/(4x),
\quad
\Im h_2 = 0,
\quad
2xh_2'+(4-\Re h_1)h_2-k=0,
\\
&
4x^2(\Re h_1)' - 2x(\Re h_1-6)\Re h_1 - 1/(8x)=0,
\\
&
a=-1/2,
\quad
p = -4,
\quad
n=-1 ;
\end{aligned}
\end{equation}
\begin{equation}
\begin{aligned}\label{scal-solh-q}
&
\Im h_1 = -1/(4x),
\quad
\Im h_2 = 0,
\quad
2xh_2'+(4-\Re h_1)h_2-k=0,
\\
&
4x^2(\Re h_1)'' + 4x(5 - \Re h_1)(\Re h_1)'-2(\Re h_1)^2+12\Re h_1+1/(8x^2)=0,
\\
&
a=-1/2,
\quad
p = -4,
\quad
n=-1 .
\end{aligned}
\end{equation}

The ODEs in equation \eqref{scal-solh-r}
can be solved in terms of Bessel functions by the following steps.
First, the Riccati transformation $\Re h_1 = -2x^2(x^{-1}f)'/f$
converts the first-order nonlinear ODE for $\Re h_1$ into the second-order linear ODE
\begin{equation}\label{scal-solh-r-riccati-f}
x^2 f'' + 2x f' - (2-x^{-2}/64)f = 0 . 
\end{equation}
A change of variables $x=1/(8z)$ and $f(x)=z^{1/2}\tilde{f}(z)$
transforms this ODE \eqref{scal-solh-r-riccati-f} into Bessel's equation
\begin{equation}
z^2\tilde{f}'' + z\tilde{f}' + (z^2-9/4)\tilde{f} = 0
\end{equation}
whose general solution is a linear combination of Bessel functions
$J_{3/2}$ and $Y_{3/2}$.
Then the general solution for ODE \eqref{scal-solh-r-riccati-f} is given by
\begin{equation}\label{scal-solh-r-riccati-f-sol}
f = (8x)^{-1/2}(C_1 J_{3/2}(1/(8x)) + C_2 Y_{3/2}(1/(8x))) . 
\end{equation}
This yields the general solution for $\Re h_1(x)$,
\begin{equation}\label{scal-solh-r-riccati-h1-f}
\Re h_1
= (4x)^{-1}\frac{C_1 J_{1/2}(1/(8x)) + C_2 Y_{1/2}(1/(8x))}{C_1 J_{3/2}(1/(8x)) + C_2 Y_{3/2}(1/(8x))} . 
\end{equation}
Next, the remaining ODE for $h_2(x)$ in equation \eqref{scal-solh-r}
becomes
\begin{equation}\label{scal-solh-r-riccati-h2-f-ode}
(x f h_2)' = (k/2)f
\end{equation}
which can be directly integrated to get
\begin{equation}\label{scal-solh-r-riccati-h2-f}
h_2 =
(k/4)\left(1-\frac{C_1 \Si(1/(8x)) - C_2 \Ci(1/(8x)) + C_3}{2\sqrt{\pi x}(C_1 J_{3/2}(1/(8x)) + C_2 Y_{3/2}(1/(8x)))}
\right)
\end{equation}
in terms of the Sine integral $\Si(x)$ and Cosine integral $\Ci(x)$ 
\cite{AbrSte}.

Similarly,
the ODEs in equation \eqref{scal-solh-q}
can be solved in terms of Coulomb functions by the following steps.
First, the second-order ODE for $\Re h_1(x)$ reduces by direct integration
to a first-order ODE
\begin{equation}\label{scal-solh-q-h1-ode}
4x^2(\Re h_1)' - 2x(\Re h_1-6)\Re h_1 - 1/(8x) +C_1=0
\end{equation}
with $C_1\neq0$.
(Note the case $C_1=0$ is covered by equation \eqref{scal-solh-r}.)
This ODE \eqref{scal-solh-q-h1-ode} is a Riccati equation
which can be converted into a Coulomb wave equation
by the transformations $\Re h_1 = 2(zf(z))'/f(z)$ and $z=1/(8x)$,
giving
\begin{equation}
f'' + (1  -C_1/z - 2/z^2) f = 0 . 
\end{equation}
The general solution is given by
\begin{equation}\label{scal-solh-q-riccati-f-whittalker}
f = C_2 F_{1}(C_1/2,z) + C_3 G_{1}(C_1/2,z)
\end{equation}
in terms of the regular and irregular Coulomb wave functions $F_L$ and $G_L$
\cite{AbrSte}. 
Hence
\begin{equation}\label{scal-solh-q-riccati-h1-f}
\Re h_1
= -C_1/(8x) + \frac{\sqrt{4+C_1^2}( C_2 F_{0}(C_1/2,1/(8x)) + C_3 G_{0}(C_1/2,1/(8x)) )}{8x( C_2 F_{1}(C_1/2,1/(8x)) + C_3 G_{1}(C_1/2,1/(8x)) )}
\end{equation}
yields the general solution for $\Re h_1(x)$.
Next, the ODE for $h_2(x)$ remaining in equation \eqref{scal-solh-q}
becomes
\begin{equation}\label{scal-solh-q-riccati-h2-f-ode}
(x f h_2)' = (k/2)f . 
\end{equation}
By directly integrating this ODE, we obtain
\begin{equation}\label{scal-solh-q-riccati-h2-f}
h_2 =
\frac{k\int^{x}_{C_4}( C_2 F_{1}(C_1/2,1/(8\xi)) + C_3 G_{1}(C_1/2,1/(8\xi)) )d\xi}
{2x( C_2 F_{1}(C_1/2,1/(8x)) + C_3 G_{1}(C_1/2,1/(8x)) )} . 
\end{equation}

Altogether from equations \eqref{scal-sol-a}--\eqref{scal-solh-q}
we obtain 17 solutions for $(h_1(x),h_2(x))$. 
We now list the resulting solutions for $(H,G)$. 

\begin{prop}\label{scal-HG-solns}
For $p\neq0$ and $n\neq 1$,
the ansatz \eqref{Hansatz}--\eqref{Gansatz} yields 17
phase-equivariant solutions of the scaling-group resolving system \eqref{scal-GH}:
{\allowdisplaybreaks
\begin{align}
&\label{scal-solGH-a}
\begin{aligned}
H = & 0,
\quad
G = -\i k|v|^p v ;
\end{aligned}\\
& \label{scal-solGH-b}
\begin{aligned}
H = & -(\i/2)x^{-1} v,
\quad
G = \left(-(n/2)x^{-1} + (\i/4)x^{-2} - \i k|v|^p\right)v ;
\end{aligned}\\
&\label{scal-solGH-c}
\begin{aligned}
H = & \i C_1 |v|^{2/n}v,
\quad
G = \left(\i {C_1}^2 |v|^{4/n} + (C_1 n-\i k)|v|^{2/n}\right)v , 
\\&
p=2/n,
\quad
n\neq 0 ;
\end{aligned}\\
&\label{scal-solGH-h}
\begin{aligned}
H = & \left(2-n \mp \i \sqrt{2k(1-2/n)} |v|^{p/2}\right)v,
\\
G = & \left(\i k(1-4/n) |v|^{p} \mp \sqrt{2k(1-2/n)} (4-n) |v|^{p/2}\right)v , 
\\&
p=2/(2-n),
\quad
n(n-2)/k>0, 
\quad
n\neq 2 ;
\end{aligned}\\
&\label{scal-solGH-f}
\begin{aligned}
H = & \left(2-n \mp \sqrt{k(2-n)} |v|^{p/2}\right)v,
\quad
G = 0 , 
\\&
p=2(3-n)/(n-2),
\quad
k(2-n)>0,
\quad
n\neq 2,3 ;
\end{aligned}\\
&\label{scal-solGH-i}
\begin{aligned}
H = & \left(2-n \mp \i\sqrt{k} |v|^{p/2}\right)v,
\quad
G = 0 , 
\\&
p=2(3-n)/(n-2),
\quad
k>0,
\quad
n\neq 2,3 ;
\end{aligned}\\
&\label{scal-solGH-d}
\begin{aligned}
H = & \left(2-n - (k/2) |v|^{-1}\right)v,
\quad
G = 0 , 
\\&
p=-1 ;
\end{aligned}\\
&\label{scal-solGH-e}
\begin{aligned}
H = & - (k/n) |v|^{-1}v,
\quad
G = 0 , 
\\&
p=-1,
\quad
n\neq 0,2 ;
\end{aligned}\\
&\label{scal-solGH-k}
\begin{aligned}
H = & \left(-1 -\i kx |v|^{-1}\right)v,
\quad
G = \left(\i k^2x^2|v|^{-2} - \i k|v|^{-1}\right)v, 
\\&
p=-1,
\quad
n=3 ;
\end{aligned}\\
&\label{scal-solGH-l}
\begin{aligned}
H = & \left(-(\i/2)x^{-1} + (k/4) |v|^{-1}\right)v,
\quad
G = \left((\i/4)x^{-2}+2x^{-1} -(k/4)x^{-1}|v|^{-1}\right)v,
\\&
p=-1,
\quad
n=-4 ;
\end{aligned}\\
&\label{scal-solGH-s}
\begin{aligned}
H = & \left(6 - \i/(2x) - (k/2)|v|^{-1}\right)v,
\quad
G = \left(-4x^{-1} + (\i/4)x^{-2} + (k/2)x^{-1}|v|^{-1}\right)v,
\\&
p = -1,
\quad
n=-4 ;
\end{aligned}\\
&\label{scal-solGH-m}
\begin{aligned}
H = & \left(2-n-(\i/2)x^{-1} \pm \sqrt{-kn/(n+2)}|v|^{2/n}\right)v,
\\
G = & \left(-(n+4)/(2n+4)+\i/(4x))x^{-1} \mp \sqrt{-kn/(n+2)}x^{-1}|v|^{2/n}\right)v,
\\&
p = 4/n,
\quad
n^2-n-4=0 ,
\quad
kn<0 ;
\end{aligned}\\
&\label{scal-solGH-o}
\begin{aligned}
H = & \left(2-n - (\i/2)x^{-1} \pm \i \sqrt{k}|v|^{2/n}\right)v,
\\
G = & \left((-4/(n+3) + \i/(4x))x^{-1} \mp \i\sqrt{k}x^{-1}|v|^{2/n}\right)v,
\\&
p = 4/n,
\quad
n^2-n-4=0,
\quad
k>0 ;
\end{aligned}\\
&\label{scal-solGH-n}
\begin{aligned}
H = & \left(2/3 - (\i/2)x^{-1} \pm \i\sqrt{-k}|v|^{3/2}\right)v,
\\
G = & \left(-(8/3)x^{-1} + (\i/4)x^{-2} \mp \sqrt{-k}(\i/x-8/3)|v|^{3/2}-2\i k|v|^3\right)v,
\\&
p = 3,
\quad
n=4/3,
\quad
k<0 ;
\end{aligned}\\
&\label{scal-solGH-p}
\begin{aligned}
H = & \left(-1 - (\i/2)x^{-1} - \i(2k/5)x|v|^{-1}\right)v,
\\
G = & \left(-(1/2)x^{-1}+(\i/4)x^{-2} +\i(4k^2/25)x^2|v|^{-2} - \i(3k/5)|v|^{-1}\right)v,
\\&
p=-1,
\quad
n=3 ;
\end{aligned}\\
&\label{scal-solGH-r}
\begin{aligned}
H = &
\bigg(
\frac{C_1\sin(1/(8x)) - C_2\cos(1/(8x)) -(k/8)(C_1\Si(1/(8x)) - C_2\Ci(1/(8x)) + C_3)|v|^{-1}}{4x\big( (8xC_1-C_2)\sin(1/(8x)) - (8xC_2+C_1)\cos(1/(8x)) \big)}
\\&\quad
-\i/(4x) +(k/4) |v|^{-1}
\bigg)v, 
\\
G = &
\bigg(
\frac{C_1\sin(1/(8x)) - C_2\cos(1/(8x)) -(k/8)(C_1\Si(1/(8x)) - C_2\Ci(1/(8x)) + C_3)|v|^{-1}}{8x^2\big( (8xC_1-C_2)\sin(1/(8x)) - (8xC_2+C_1)\cos(1/(8x)) \big)}
\\&\quad
+ 1/x +\i/(8x^2) - (k/(8x)) |v|^{-1}
\bigg)v,
\\&
p=-1,
\quad
n=-4 ;
\end{aligned}\\
&\label{scal-solGH-q}
\begin{aligned}
H = &
\bigg((-2\i - C_1)/(8x) + \frac{\sqrt{4+C_1^2}( C_2 F_{0}(C_1/2,1/(8x)) + C_3 G_{0}(C_1/2,1/(8x)) )}{8x( C_2 F_{1}(C_1/2,1/(8x)) + C_3 G_{1}(C_1/2,1/(8x)) )}
\\&
+ \frac{k\int^{x}_{C_4}( C_2 F_{1}(C_1/2,1/(8\xi)) + C_3 G_{1}(C_1/2,1/(8\xi)) )d\xi}
{2x( C_2 F_{1}(C_1/2,1/(8x)) + C_3 G_{1}(C_1/2,1/(8x)) )}
|v|^{-1}\bigg)v,
\\
G = &
\bigg((1-\i C_1/2)(1+\i/(8x))/x - \frac{\sqrt{4+C_1^2}( C_2 F_{0}(C_1/2,1/(8x)) + C_3 G_{0}(C_1/2,1/(8x)) )}{16x^2( C_2 F_{1}(C_1/2,1/(8x)) + C_3 G_{1}(C_1/2,1/(8x)) )}
\\&
- \frac{k\int^{x}_{C_4}( C_2 F_{1}(C_1/2,1/(8\xi)) + C_3 G_{1}(C_1/2,1/(8\xi)) )d\xi}
{4x^2( C_2 F_{1}(C_1/2,1/(8x)) + C_3 G_{1}(C_1/2,1/(8x)) )}
|v|^{-1}\bigg)v,
\\&
p=-1,
\quad
n=-4 . 
\end{aligned}
\end{align}
\endallowdisplaybreaks}
None of these solutions
satisfy the scaling-invariance condition \eqref{scal-inv-GHsoln}.
\end{prop}

By means of the invariants $x=t/r^2$, $v=r^{2/p}u$, $\bar{v}=r^{2/p}\bar{u}$
and the canonical coordinate $y=(1/2)\ln t$ of $\X_{\scal}$,
we can write the differential invariants \eqref{scal-diffinv} of $\X_{\scal}$
in terms of $x,y$-derivatives
\begin{equation}
G=r^2 D_t v = v_x + (2x)^{-1}v_y,
\quad
H=rD_r v -(2/p)v = -2x v_x -(2/p)v . 
\end{equation}
Hence, each phase-equivariant solution $(G=g(x,|v|)v,H=h(x,|v|)v)$ of
the scaling-group resolving system \eqref{scal-GH}
yields a pair of DEs \eqref{scal-odes}
given by
\begin{subequations}
\begin{align}
& v_y =  2xG +H +(2/p)v =( 2x g(x,|v|) + h(x,|v|) +2/p )v , 
\\
& v_x= -(2x)^{-1}(H+(2/p)v)=-(2x)^{-1}( h(x,|v|)+ 2/p )v . 
\end{align}
\end{subequations}
These DEs determine a two-parameter family of solutions
$u=f(t,r,c_1)\exp(\i c_2)$ of the radial gNLS equation \eqref{nls},
corresponding to orbits of the two-dimensional symmetry group 
$\G$ generated by $\X_{\scal}$ and $\X_{\phas}$.
In polar form $u=r^{-2/p} A\exp(\i \Phi)$,
the solution families are given by
the line integral formula \eqref{y-soln}--\eqref{Phi-soln}
in the case $\Re\hat g\neq 0$
and the integration formula \eqref{Phi-soln'}--\eqref{A-soln'}
in the case $\Re\hat g=0$,
using the notation $\hat g=2xg+h+2/p$ and $\hat h=-(2x)^{-1}(h+2/p)$.
This establishes a group-invariant mapping from
phase-equivariant solutions of the scaling-group resolving system \eqref{scal-GH}
into a class of solutions of the radial gNLS equation \eqref{nls} 
satisfying the scaling invariance property \eqref{scal-orbit},
with the integration constants given by expressions \eqref{c1c2-case1}--\eqref{c1c2-case2} 
in terms of the group parameter $\lambda=\exp(\epsilon)$.
An inverse mapping can be constructed by the same steps
explained for the time-translation-group resolving system.

This completes the proof of \lemref{scal-map}. 
The proof of \lemref{similarity-condition} corresponds to 
the integration case $\hat g=0$.

\subsection{Results for the inversion-group resolving system}
\label{inver-GH-solns}

The overdetermined systems of algebraic-differential equations
arising from reduction of the inversion-group resolving system \eqref{inver-GH}
under the separation of variables ansatz \eqref{Hansatz}--\eqref{Gansatz}
with $p=4/n$
admit non-zero solutions $(h_1(x),h_2(x))$ 
only in the cases $a=2/n$, $a=1/n$, $a\neq-2/n,1/n,2/n,2/(3n)$.
For $n\neq0,1$, the solutions are given by:
\begin{equation}\label{inver-sol-a}
h_1=h_2=0 ;
\end{equation}
\begin{equation}\label{inver-sol-b}
\begin{aligned}
& 
h_1=\Re h_2=0,
\quad
(x h_2)'=0,
\\
& 
a=1/n ;
\end{aligned}
\end{equation}
\begin{equation}\label{inver-sol-c}
\begin{aligned}
&
h_1=2-n,
\quad
\Im h_2=0,
\quad
h_2{}^2=-kn/(n+2),
\\
&
a=1/n,
\quad
n^2-n-4=0,
\quad
kn<0 ;
\end{aligned}
\end{equation}
\begin{equation}\label{inver-sol-a1}
\begin{aligned}
&
h_1=2-n,
\quad
\Re h_2=0,
\quad
h_2{}^2=-k,
\\
&
a=1/n,
\quad
n^2-n-4=0,
\quad
k>0 ;
\end{aligned}
\end{equation}
\begin{equation}\label{inver-sol-a2}
\begin{aligned}
&
h_1=2/3,
\quad
\Re h_2=0,
\quad
h_2{}^2=k,
\\
& 
a=1/n,
\quad
n=4/3,
\quad
k<0 ;
\end{aligned}
\end{equation}
\begin{equation}\label{inver-sol-e-case1-subcase}
\begin{aligned}
&
h_1=\Im h_2=0,
\quad
xh_2'+4h_2-k=0,
\\
&
a=-1/2,
\quad
n=-4 ;
\end{aligned}
\end{equation}
\begin{equation}
\begin{aligned}\label{inver-sol-e-case1-generalcase}
& 
\Im h_1=\Im h_2=0,
\quad
xh_1'-h_1^2+6h_1=0,
\quad
xh_2'+(4-h_1)h_2-k=0,
\\
&
a=-1/2,
\quad
n=-4 ;
\end{aligned}
\end{equation}
\begin{equation}
\begin{aligned}\label{inver-sol-e-case2}
& 
\Im h_1=\Im h_2=0,
\quad
x^2h_1''-x(2h_1-9)h_1'-2h_1(h_1-6)=0,
\\&
xh_2'+(4-h_1)h_2-k=0,
\\
&
a=-1/2,
\quad
n=-4 . 
\end{aligned}
\end{equation}

The ODEs in equations
\eqref{inver-sol-b}, \eqref{inver-sol-e-case1-subcase}, \eqref{inver-sol-e-case1-generalcase}
are simple to solve,
while the ODEs in equation \eqref{inver-sol-e-case2}
can be solved in terms of Bessel functions by the same steps used to solve
the similar ODEs in equation \eqref{trans-sol-d-case2}.
Altogether from equations \eqref{inver-sol-a}--\eqref{inver-sol-e-case2}
we obtain 9 solutions for $(h_1(x),h_2(x))$. 
We now list the resulting solutions for $(H,G)$. 

\begin{prop}\label{inver-HG-solns}
For $n\neq0,1$, the ansatz \eqref{Hansatz}--\eqref{Gansatz} yields 
9 phase-equivariant solutions of 
the inversion-group resolving system \eqref{inver-GH}
with $p=4/n$:
{\allowdisplaybreaks
\begin{align}
& \label{inver-solGH-a}
\begin{aligned}
H = &0,
\quad
G=-\i k |v|^{4/n}v ;
\end{aligned}\\
& \label{inver-solGH-b}
\begin{aligned}
H = &\i C_1x^{-1}|v|^{2/n}v,
\quad
G = \left(\i C_1^2x^{-2}|v|^{4/n} +C_1nx^{-1}|v|^{2/n}-\i k |v|^{4/n}\right)v,
\\&
C_1\neq 0 ;
\end{aligned}\\
& \label{inver-solGH-a2}
\begin{aligned}
H = & \left(2/3 \pm \i\sqrt{-k}|v|^{3/2}\right)v,
\quad
G= \left(\pm(8/3)\sqrt{-k} |v|^{3/2} -\i2k|v|^{3}\right)v,
\\&
n=4/3,
\quad
k<0 ;
\end{aligned}\\
& \label{inver-solGH-c}
\begin{aligned}
H = & \left(2-n \pm \sqrt{-kn/(n+2)}|v|^{2/n}\right)v,
\quad
G= 0, 
\\&
n^2-n-4=0,
\quad
kn<0 ;
\end{aligned}\\
& \label{inver-solGH-a1}
\begin{aligned}
H = & \left(2-n \pm \i\sqrt{k}|v|^{2/n}\right)v,
\quad
G=0,
\\&
n^2-n-4=0,
\quad
k>0 ;
\end{aligned}\\
& \label{inver-solGH-e-1}
\begin{aligned}
H = &(C_1x^{-4} +k/4)|v|^{-1}v,
\quad
G=0,
\\&
n=-4 ;
\end{aligned}\\
& \label{inver-solGH-e-2}
\begin{aligned}
H = & \left((1+C_1x^6)^{-1}(6 + (k/4)(C_2x^2-3)|v|^{-1})+(k/4)|v|^{-1}\right)v,
\quad
G=0,
\\&
n=-4 ;
\end{aligned}\\
& \label{inver-solGH-e-3}
\begin{aligned}
H = &
\bigg(\sqrt{C_1}\left(x(C_2J_3(\sqrt{C_1}/x) + C_3 Y_3(\sqrt{C_1}/x))\right)^{-1}
\times\\&\qquad
\left((C_2J_2(\sqrt{C_1}/x) + C_3 Y_2(\sqrt{C_1}/x))(1+(k/C_1)x^2|v|^{-1})+C_4|v|^{-1}\right)\bigg)v,
\\
G = &
\i C_1 x^{-2}v,
\\&
n=-4,
\quad
C_1>0 ;
\end{aligned}\\
& \label{inver-solGH-e-4}
\begin{aligned}
H = &
\bigg(\sqrt{C_1}\left(x(C_2I_3(\sqrt{C_1}/x) + C_3 e^{\i 3\pi} K_3(\sqrt{C_1}/x))\right)^{-1}
\times\\&\qquad
\left((C_2I_2(\sqrt{C_1}/x) + C_3 e^{\i 2\pi} K_2(\sqrt{C_1}/x))(1-(k/C_1)x^2|v|^{-1})+C_4|v|^{-1}\right)\bigg)v,
\\
G = &
-\i C_1x^{-2}v,
\\&
n=-4,
\quad
C_1>0 .
\end{aligned}
\end{align}
\endallowdisplaybreaks}
Only solutions \eqref{inver-solGH-c}--\eqref{inver-solGH-e-2}
satisfy the pseudo-conformal-invariance condition \eqref{inver-inv-GHsoln}.
\end{prop}

The invariants \eqref{inver-inv}
and the canonical coordinate $y=-1/t$ of $\X_{\inver}$
can be used to write
the differential invariants \eqref{inver-diffinv} of $\X_{\inver}$
in the form of $x,y$-derivatives
\begin{equation}
G=r^2( D_t v +(r/t)D_r v )= x^{-2} v_y,
\quad
H=rD_r v -(n/2)v = -x v_x -(n/2)v .
\end{equation}
Then each solution $(G=g(x,|v|)v,H=h(x,|v|)v)$ of
the inversion-group resolving system \eqref{inver-GH}
yields a pair of DEs \eqref{inver-odes}
given by
\begin{equation}
v_y =  x^2 G =x^2 g(x,|v|) v,
\quad
v_x= -x^{-1}(H+(n/2)v)=-x^{-1}( h(x,|v|)+ n/2 )v
\end{equation}
which determines a two-parameter family of solutions
$u=f(t,r,c_1)\exp(\i c_2)$ of the radial gNLS equation \eqref{nls},
corresponding to orbits of the two-dimensional symmetry group 
$\G$ generated by $\X_{\inver}$ and $\X_{\phas}$.
These solution families are given by the polar form
$u=r^{-2/p} A\exp(\i (\Phi-r^2/(4t)))$
obtained from the line integral formula \eqref{y-soln}--\eqref{Phi-soln}
in the case $\Re\hat g\neq 0$
and the integration formula \eqref{Phi-soln'}--\eqref{A-soln'}
in the case $\Re\hat g=0$,
where $\hat g=x^2 g$ and $\hat h=-x^{-1}(h+n/2)$.
This establishes a group-invariant mapping from
phase-equivariant solutions of the inversion-group resolving system \eqref{inver-GH}
into solutions of the radial gNLS equation \eqref{nls} satisfying
the pseudo-conformal invariance property \eqref{inver-orbit}
such that the relations \eqref{c1c2-case1}--\eqref{c1c2-case2} hold.
An inverse mapping can be constructed using the same steps
explained for the time-translation-group resolving system.

This completes the proof of \lemref{inver-map},
while the proof of \lemref{conformal-condition} corresponds to 
the integration case $\hat g=0$.

\section{Main results}
\label{results}

Here we will write out all of the radial gNLS solutions $u(t,r)$
arising from Propositions~\ref{trans-HG-solns}, \ref{scal-HG-solns}, and \ref{inver-HG-solns},
via the quadrature formulas
\eqref{y-soln}--\eqref{Phi-soln} and \eqref{Phi-soln'}--\eqref{A-soln'}.

\begin{thm}\label{u-solns}
The radial gNLS equation \eqref{nls} has the following exact solutions
arising from the explicit solutions of the group resolving systems
\eqref{trans-GH}, \eqref{scal-GH}, \eqref{inver-GH} for $n\neq 1$:
{\allowdisplaybreaks
\begin{align}
&\label{trans-rnls-sol1}
\begin{aligned}
u = &
(c_2/k)^{1/p}\exp(\i c_1 - \i c_2 t) ;
\end{aligned}\\
&\label{trans-rnls-sol2}
\begin{aligned}
u = &
(c_2+c_3t)^{-n/2}\exp\Big(\i c_1 - \frac{\i c_3r^2}{4(c_2+c_3t)}
+ \frac{2\i k}{c_3(np-2)}(c_2+c_3t)^{1-np/2}\Big), 
\\&
p\neq 2/n,
\quad
n\neq0,
\quad
c_3\neq0 ;
\end{aligned}\\
&\label{trans-rnls-sol3}
\begin{aligned}
u = &
(c_2+ c_3t)^{-n/2}\exp\Big(\i c_1 - \frac{\i c_3 r^2}{4(c_2+c_3t)} - \frac{\i k}{c_3}\ln |c_2+c_3t|\Big), 
\\&
p= 2/n,
\quad
n\neq0,
\quad
c_3\neq0 ;
\end{aligned}\\
&\label{scal-rnls-sol2}
\begin{aligned}
u = &
(\pm\sqrt{n(n-2)/(2k)})^{2-n}\big((c_2 +(n-4)t)/r\big)^{n-2}
\exp\big(\i c_1 +\i(1-n/2) r^2/(c_2 + (n-4)t)\big), 
\\&
p=2/(2-n),
\quad
n(n-2)/k>0,
\quad
n\neq 2 ;
\end{aligned}\\
&\label{trans-rnls-sol4}
\begin{aligned}
u = &
\big(k(n-3)^2/(2-n)^3\big)^{(2-n)/(6-2n)}\big( r + c_2r^{3-n}\big)^{(2-n)/(3-n)}\exp(\i c_1), 
\\&
p=2(3-n)/(n-2),
\quad
k(2-n)>0,
\quad
n\neq 2,3 ;
\end{aligned}\\
& \label{scal-rnls-sol3}
\begin{aligned}
u = &
\big(c_2^2(n-2)^2/k\big)^{(n-2)/(6-2n)}r^{2-n}\exp(\i c_1 + \i c_2r^{n-2}), 
\\&
p=2(3-n)/(n-2),
\quad
k>0,
\quad
n\neq 2,3,
\quad
c_2\neq 0 ;
\end{aligned}\\
& \label{trans-rnls-sol8}
\begin{aligned}
u = &
\bigg(-k/c_6 + r^{1-n/2}
\big(c_2J_{|1-n/2|}(\sqrt{c_6}r) + c_3 Y_{|1-n/2|}(\sqrt{c_6}r)\big)
\\&\quad
\times\Big( 1+c_5 \int_{c_4}^{r} z^{-1}(c_2J_{|1-n/2|}(\sqrt{c_6}z) + c_3 Y_{|1-n/2|}(\sqrt{c_6}z))^{-2}\;dz \Big)
\bigg)\exp(\i c_1 +\i c_6t), 
\\&
p=-1,
\quad
c_6>0 ;
\end{aligned}\\
&\label{trans-rnls-sol9}
\begin{aligned}
u = &
\bigg(k/c_6 + r^{1-n/2}
\big(c_2I_{|1-n/2|}(\sqrt{c_6}r) + c_3 K_{|1-n/2|}(\sqrt{c_6}r)\big)
\\&\quad
\times\Big( 1+c_5 \int_{c_4}^{r} z^{-1}(c_2I_{|1-n/2|}(\sqrt{c_6}z) + c_3 K_{|1-n/2|}(\sqrt{c_6}z))^{-2}\;dz \Big)
\bigg)\exp(\i c_1 - \i c_6t), 
\\&
p=-1,
\quad
c_6>0 ;
\end{aligned}\\
& \label{trans-rnls-sol5}
\begin{aligned}
u = &
(-kr^2/(2n) + c_3r^{2-n} + c_2)\exp(\i c_1), 
\\&
p=-1,
\quad
n\neq 0,2 ;
\end{aligned}\\
&\label{scal-rnls-sol7}
\begin{aligned}
u = &
(c_2/(rt^{1/2}))
\exp\big(\i c_1 - \i r^2/(4t) - 2\i krt^{3/2}/(5c_2) + \i k^2t^4/(25c_2^2)\big), 
\\&
p=-1,
\quad
n=3 ;
\end{aligned}\\
&\label{scal-rnls-sol4}
\begin{aligned}
u = &
(c_2/r)\exp\big(\i c_1 - \i ktr/c_2 + \i k^2t^3/(3c_2^2)\big), 
\\&
p=-1,
\quad
n=3 ;
\end{aligned}\\
&\label{trans-rnls-sol6}
\begin{aligned}
u = &
(-kr^2/4 + c_3\ln r + c_2)\exp(\i c_1), 
\\&
p=-1,
\quad
n=2 ;
\end{aligned}\\
&\label{trans-rnls-sol7}
\begin{aligned}
u = &
(-(k/2)r^2\ln r + c_3 r^2 + c_2)\exp(\i c_1), 
\\&
p=-1,
\quad
n=0 ;
\end{aligned}\\
&\label{inver-rnls-sol4}
\begin{aligned}
u = &
\big((k/8)r^2 + c_3r^6/t^4 + c_2 t^2\big)\exp(\i c_1 - \i r^2/(4t)), 
\\&
p=-1,
\quad
n=-4 ;
\end{aligned}\\
&\label{inver-rnls-sol5}
\begin{aligned}
u = &
\bigg(-(k/c_6)t^2 + (r^3/t)\left(c_2J_3(\sqrt{c_6}r/t) + c_3 Y_3(\sqrt{c_6}r/t)\right)
\\&\quad
\times\Big( 1+c_5 \int_{c_4}^{r/t} z^{-1}(c_2J_3(\sqrt{c_6}z) + c_3 Y_3(\sqrt{c_6}z))^{-2} \;dz \Big)
\bigg) \exp\big(\i c_1 - \i c_6/t - \i r^2/(4t)\big), 
\\&
p=-1,
\quad
n=-4,
\quad
c_6>0 ;
\end{aligned}\\
&\label{inver-rnls-sol6}
\begin{aligned}
u = &
\bigg((k/c_6)t^2 + (r^3/t)(c_2I_3(\sqrt{c_6}r/t) + c_3 K_3(\sqrt{c_6}r/t))
\\&\quad
\times\Big( 1+c_5 \int_{c_4}^{r/t} z^{-1}(c_2I_3(\sqrt{c_6}z) + c_3 K_3(\sqrt{c_6}z))^{-2} \;dz \Big)
\bigg) \exp\big(\i c_1 + \i c_6/t - \i r^2/(4t)\big), 
\\&
p=-1,
\quad
n=-4,
\quad
c_6>0 ;
\end{aligned}\\
&\label{inver-rnls-sol3}
\begin{aligned}
u = &
\big(\pm\sqrt{-k(1+3/n)/2}\big)^{-n/2}
\big(r + c_2t^{-1+4/n}r^{2(1-2/n)}\big)^{-n/2}
\exp(\i c_1 - \i r^2/(4t)), 
\\&
p=8/(1\pm\sqrt{17}) = (\pm\sqrt{17}-1)/2,
\quad
n=(1\pm\sqrt{17})/2,
\quad
kn<0 ;
\end{aligned}\\
&\label{scal-rnls-sol6}
\begin{aligned}
u = &
\big(c_2^2(8-3n)/k\big)^{n/4}r^{2-n}t^{-2+n/2}
\exp\big(\i c_1 - \i r^2/(4t) + \i c_2r^{n-2}t^{2-n}\big), 
\\&
p=8/(1\pm\sqrt{17}) = (\pm\sqrt{17}-1)/2,
\quad
n=(1\pm\sqrt{17})/2,
\quad
k>0 ;
\end{aligned}\\
&\label{scal-rnls-sol5}
\begin{aligned}
u = &
(-16k)^{-1/3}r^{2/3} (t(1 + c_2t))^{-2/3}
\exp\big(\i c_1 - \i r^2(1 + 2c_2t)/(8t(1 + c_2t))\big), 
\\&
p = 3,
\quad
n=4/3,
\quad
k<0 ;
\end{aligned}\\
&\label{scal-rnls-sol-r}
\begin{aligned}
u = &
(k/8)
\big((c_2r^2+8c_3t)\cos(r^2/(8t))+(c_3r^2-8c_2t)\sin(r^2/(8t))\big)
\\&\quad \times
\int^{r^2/(8t)}_{c_4}
\frac{\xi^2(c_2\Si(\xi)-c_3\Ci(\xi))+(c_3\xi-c_2)\sin(\xi)+(c_2\xi+c_3)\cos(\xi)}{\big( (c_2-c_3\xi)\sin(\xi)-(c_3+c_2\xi)\cos(\xi)\big)^2}d\xi
\\&\quad \times
\exp\big(\i c_1 - \i r^2/(8t)\big), 
\\&
p = -1,
\quad
n=-4 ;
\end{aligned}\\
&\label{scal-rnls-sol-q}
\begin{aligned}
u = &
-(kr^2/4)\left(c_3F_1(c_2, r^2/(8t)) + c_4G_1(c_2, r^2/(8t))\right)
\\&\qquad \times
\int^{r^2/(8t)}_{c_5}
\frac{\big( c_3Fi_1(c_2, \xi) + c_4Gi_1(c_2, \xi) \big)}
{\big( c_3F_1(c_2, \xi) + c_4G_1(c_2, \xi) \big)^2}d\xi
\exp(\i c_1 - \i r^2/(8t) - \i c_2\ln t), 
\\&
p = -1,
\quad
n=-4,
\quad
c_2\neq0 , 
\end{aligned}
\end{align}
\endallowdisplaybreaks}
where 
\begin{equation*}
Fi_L(\rho, \xi)=\int^{\xi}_{c_6}z^{-2}F_L(\rho, z)dz,
\quad
Gi_L(\rho, \xi)=\int^{\xi}_{c_6}z^{-2}G_L(\rho, z)dz,
\quad
c_6\neq0 . 
\end{equation*}
\end{thm}
Here $\Si(x)$ and $\Ci(x)$ denote the Sine integral and Cosine integral;
$F_L(\rho,x)$ and $G_L(\rho,x)$ denote the regular and irregular Coulomb wave functions.
(See \Ref{AbrSte}.) 

\begin{rem}
Solutions 
\eqref{trans-rnls-sol2}, 
\eqref{trans-rnls-sol8}--\eqref{trans-rnls-sol5}, 
\eqref{trans-rnls-sol6}, 
\eqref{trans-rnls-sol7} 
come from the time-translation-group resolving system \eqref{trans-GH}.
Solutions
\eqref{scal-rnls-sol7}, \eqref{scal-rnls-sol4}, 
\eqref{scal-rnls-sol-r}, \eqref{scal-rnls-sol-q}
come from the scaling-group resolving system \eqref{scal-GH}.
Solutions 
\eqref{inver-rnls-sol4}--\eqref{inver-rnls-sol6}
come from the inversion-group resolving system \eqref{inver-GH}.
Of the remaining solutions, 
\eqref{trans-rnls-sol1} and 
\eqref{trans-rnls-sol3}--\eqref{scal-rnls-sol3}
come from both the time-translation-group and scaling-group resolving systems,
while 
\eqref{inver-rnls-sol3}--\eqref{scal-rnls-sol5}
come from both the inversion-group and scaling-group resolving systems.
\end{rem}

The full group of point symmetries \eqref{phas-group}--\eqref{inver-group}
for the radial gNLS equation \eqref{nls}
can be applied to each of the solutions $u=f(t,r)$ listed in \thmref{u-solns}.
Phase rotations \eqref{phas-group} and scalings \eqref{scal-group}
change only the constants appearing in these solutions,
while time-translations \eqref{trans-group} at most shift $t$ by a new constant.
In contrast, inversions \eqref{inver-group} have a non-trivial action
on solutions, which is summarized as follows:
\eqref{trans-rnls-sol1} with $p=4/n$
is transformed to the $p=4/n$ case of \eqref{trans-rnls-sol2}
up to phase shift (via $t/(1+c_3t)=(1/c_3)-(1/c_3)/(1+c_3t)$);
\eqref{trans-rnls-sol2} with $p=4/n$
is unchanged up to phase shift;
\eqref{trans-rnls-sol4} with $p=4/n=2(3-n)/(n-2)$
is transformed to \eqref{inver-rnls-sol3} up to time-translation;
\eqref{scal-rnls-sol3} with $p=4/n=2(3-n)/(n-2)$
is transformed to \eqref{scal-rnls-sol6} up to time-translation;
\eqref{trans-rnls-sol5} with $n=-4$
is transformed to \eqref{inver-rnls-sol4} up to time-translation;
\eqref{trans-rnls-sol8} with $n=-4$
is transformed to \eqref{inver-rnls-sol5} up to time-translation;
\eqref{trans-rnls-sol9} with $n=-4$
is transformed to \eqref{inver-rnls-sol6} up to time-translation;
\eqref{inver-rnls-sol4}--\eqref{scal-rnls-sol6}
are unchanged;
\eqref{scal-rnls-sol5}
is transformed to
\begin{equation}
\label{scal-rnls-sol5-apply-inver}
\begin{aligned}
u = &
(-16k)^{-1/3}r^{2/3} (t(1+(c_2+c_3)t))^{-2/3}
\\&\qquad
\times
\exp\big(\i c_1 - \i r^2(1+2(c_2+c_3)t)/(8t(1+(c_2+c_3)t))\big),
\\&
p = 3,
\quad
n=4/3,
\quad
k<0 ;
\end{aligned}
\end{equation}
\eqref{scal-rnls-sol2} up to time-translation with $p=4/n=2/(2-n)$
is also transformed to \eqref{scal-rnls-sol5-apply-inver};
\eqref{scal-rnls-sol-r} and \eqref{scal-rnls-sol-q}
are respectively transformed to
\begin{equation}
\label{scal-rnls-sol-r-apply-inver}
\begin{aligned}
u = &
(k/8)
\big((c_2r^2+8c_3t(1+c_5t))\cos(r^2/(8t(1+c_5t)))
\\&\quad
+(c_3r^2-8c_2t(1+c_5t))\sin(r^2/(8t(1+c_5t)))\big)
\\&\quad \times
\int^{r^2/(8t(1+c_5t))}_{c_4}
\frac{\xi^2(c_2\Si(\xi)-c_3\Ci(\xi))+(c_3\xi-c_2)\sin(\xi)+(c_2\xi+c_3)\cos(\xi)}{\big( (c_2-c_3\xi)\sin(\xi)-(c_3+c_2\xi)\cos(\xi)\big)^2}d\xi
\\&\quad \times
\exp\big(\i c_1 - \i r^2/(8t(1+c_5t)) - \i c_5r^2/(4(1+c_5t))\big),
\\&
p = -1,
\quad
n=-4, 
\end{aligned}\\
\end{equation}
and
\begin{equation}
\label{scal-rnls-sol-q-apply-inver}
\begin{aligned}
u = &
-(kr^2/4)\big(c_3F_1(c_2, r^2/(8t(1+c_7t))) + c_4G_1(c_2, r^2/(8t(1+c_7t)))\big)
\\&\qquad \times
\int^{r^2/(8t(1+c_7t))}_{c_5}
\frac{c_3Fi_1(c_2, \xi) + c_4Gi_1(c_2, \xi)}
{\left(c_3F_1(c_2, \xi) + c_4G_1(c_2, \xi)\right)^2}d\xi
\\&\qquad \times
\exp\big(\i c_1 - \i r^2/(8t(1+c_7t)) - \i c_7r^2/(4(1+c_7t)) - \i c_2\ln (t/(1+c_7t))\big),
\\&
p = -1,
\quad
n=-4, 
\quad
c_2\neq0 . 
\end{aligned}
\end{equation}
These solutions \eqref{scal-rnls-sol5-apply-inver}--\eqref{scal-rnls-sol-q-apply-inver}
fall outside of the solutions listed in \thmref{u-solns}
up to time-translations, scalings, and phase shifts.

Hence we have the following result.

\begin{thm}\label{conformal-u-solns}
For $p=4/n$,
the pseudo-conformal symmetry subgroup \eqref{inver-group}
applied to the exact solutions \eqref{trans-rnls-sol1}--\eqref{scal-rnls-sol-q}
of the radial gNLS equation \eqref{nls}
yields three additional exact solutions
\eqref{scal-rnls-sol5-apply-inver}--\eqref{scal-rnls-sol-q-apply-inver}.
\end{thm}

Finally, we note that 
solutions \eqref{scal-rnls-sol-q} and \eqref{scal-rnls-sol-q-apply-inver}
do not converge if $c_6=0$ (in the integrals of the Coulomb functions), 
while solutions \eqref{scal-rnls-sol-r} and \eqref{scal-rnls-sol-r-apply-inver}
do not converge if $c_3=c_4=0$ when $c_2\neq0$.

\subsection{Analytical features}

We now discuss some basic analytical features of the solutions
in \thmref{u-solns} and \thmref{conformal-u-solns}.
Firstly, the solutions will be divided into two classes:
(I) solutions \eqref{trans-rnls-sol1}--\eqref{trans-rnls-sol6}
in which the allowed values of $n$ are positive integers;
(II) solutions \eqref{trans-rnls-sol7}--\eqref{scal-rnls-sol-q-apply-inver}
in which the allowed values of $n$ are non-positive integers or non-integers.
Class (I) describes
$n$-dimensional radial waves and monopoles of the gNLS equation \eqref{nls}, \eqref{nls-dim},
whereas class (II) is interpreted as describing
two-dimensional radial waves and monopoles of the planar gNLS equation \eqref{nls}
containing an extra point-source term $(m-1) u_r/r$ \cite{AncFen}
with a parameter $m=n-1$ (which is applicable for any value of $n\in \Rnum$).

Secondly, within each class (I) and (II),
the solutions will be categorized by their dynamical behaviour:
static, \ie/ $u=f(r)$;
time-periodic, \ie/ $u=f(r)\exp(\i \omega t)$;
dispersive, \ie/ $|u|\rightarrow 0$ for $t\rightarrow\infty$;
blow-up, \ie/ $|u|\rightarrow \infty$ for $t\rightarrow T<\infty$;
non-dispersive, \ie/ $|u|$ bounded away from $0$ for $t\rightarrow\infty$.
Additionally, the smoothness of the solutions at $r=0$ will be classified
by the conditions:
$\lim_{r\rightarrow 0}|u| < \infty$ and $\lim_{r\rightarrow 0}|u_r| =0$,
\ie/ regular;
$\lim_{r\rightarrow 0}|u| < \infty$ and $\lim_{r\rightarrow 0}|u_r| \neq 0$,
\ie/ conical;
$\lim_{r\rightarrow 0}|u| = \infty$,
\ie/ singular.

Thirdly, the invariance property of each solution
with respect to the symmetry group of the gNLS equation
will be listed.

A summary of these results is presented in
Tables~\ref{classI}, \ref{classII}, and \ref{symminv}.

\section{Concluding remarks}
\label{remarks}

Out of the 24 gNLS solutions \eqref{trans-rnls-sol1}--\eqref{scal-rnls-sol-q-apply-inver}
we have obtained in Theorems~\ref{u-solns} and~\ref{conformal-u-solns},
the time-translation invariant solutions
\eqref{trans-rnls-sol4}, \eqref{trans-rnls-sol5}, 
\eqref{trans-rnls-sol6}, \eqref{trans-rnls-sol7},
the pseudo-conformal invariant solutions 
\eqref{trans-rnls-sol2} for $p=4/n$ and \eqref{inver-rnls-sol3}
were derived in recent work \cite{AncFen}
studying group-invariant solutions of the radial gNLS equation \eqref{nls}
in multi-dimensions,
while the general non-invariant form of solution \eqref{trans-rnls-sol2}
for $p\neq 4/n$ appears in \Ref{PolZai} (without a derivation).

The remaining 18 solutions are new (to the best knowledge of the authors).
Relative to the symmetry group \eqref{phas-group}--\eqref{inver-group}
of the radial gNLS equation \eqref{nls},
15 of these new solutions are group-invariant
and the other 3 new solutions are non-invariant,
as summarized in Table~\ref{symminv}.

Altogether, these 24 solutions encompass a wide range of different
dynamical behaviours:
static; time-periodic; dispersive; blow-up; and non-dispersive.
In particular,
one case of solution \eqref{trans-rnls-sol3} exhibits a similarity blow-up
\eqref{supercrit-blowup} in which $|u|\rightarrow\infty$ in a finite time $t$
(though only for the subcritical power $p=2/n$),
and another case of this solution displays dispersion
such that $|u|\rightarrow 0$ for long times $t\rightarrow\infty$
(again for the subcritical power $p=2/n$).
Other solutions exist for special nonlinearity powers
$p=2/(2-n)$, $p=(2n-6)/(2-n)$ which are not distinguished by
the symmetry structure of the radial gNLS equation \eqref{nls}.

A detailed discussion of the interesting analytical features of all of
the solutions will be given in a forthcoming paper \cite{Anc}.

The method we have used in the present work can be applied more generally
to find explicit exact solutions to other complex ($U(1)$-invariant)
semilinear evolutions in $n\geq 1$ dimensions, such as
derivative-type gNLS equations
$\i u_t=u_{xx}  +\i( a|u|^p u_x + b (|u|^p)_x u )$
and mKdV-type equations
$u_t = u_{xxx} + a|u|^p u_x + b (|u|^p)_x u$
in one dimension,
and Landau-Ginzburg equations
$\i u_t=\triangle u +\i a u + b|u|^p u$,
Cahn-Hilliard equations
$u_t=\triangle(\triangle u +a u +b|u|^p u)$,
and Kuramoto-Sivashinsky equations
$u_t=\triangle^2 u +a\triangle u +b|\nabla u|^p u$
in multi-dimensions.

\section*{Tables}

\begin{table}[h!]
\begin{center}
\begin{tabular}{|c|c|c|c|c|c|}
\hline
solution
& $\begin{aligned} \text{nonlin.}\\\text{coeff. $k$} \end{aligned}$
& power $p\neq0$
& dimen. $n>1$
& $\begin{aligned} \text{dynamical}\\\text{behaviour} \end{aligned}$
& $\begin{aligned} \text{regularity}\\\text{at $r=0$} \end{aligned}$
\\
\hline
\hline
\eqref{trans-rnls-sol1}
& $\neq 0$
& any
& any
& time-periodic
& regular
\\
\hline
\eqref{trans-rnls-sol2}
& $\neq 0$
& $\neq 2/n$
& any
& $\begin{aligned} c_2/c_3>0, \text{ dispersive }\\c_2/c_3<0, \text{ blow-up } \end{aligned}$
& regular
\\
\hline
\eqref{trans-rnls-sol3}
& $\neq 0$
& $2/n$
& any
& $\begin{aligned} c_2/c_3>0, \text{ dispersive }\\c_2/c_3<0, \text{ blow-up } \end{aligned}$
& regular
\\
\hline
\eqref{scal-rnls-sol2}
& $>0$
& $2/(2-n)$
& $\neq 2$
& non-dispersive
& singular
\\
\hline
\eqref{trans-rnls-sol4}
& $<0$
& $(6-2n)/(n-2)$
& $\neq 2,3$
& static
& $\begin{aligned} c_2=0, \text{ regular }\\ c_2\neq0, \text{ singular } \end{aligned}$
\\
\hline
\eqref{scal-rnls-sol3}
& $>0$
& $(6-2n)/(n-2)$
& $\neq 2,3$
& static
& singular
\\
\hline
\eqref{trans-rnls-sol8}
& $\neq 0$
& $-1$
& any
& time-periodic
& $\begin{aligned} c_3=c_5=0, \text{ regular }\\ c_3\neq 0, \text{ singular }\\ c_5\neq 0, \text{ singular } \end{aligned}$
\\
\hline
\eqref{trans-rnls-sol9}
& $\neq 0$
& $-1$
& any
& time-periodic
& $\begin{aligned} c_3=c_5=0, \text{ regular }\\ c_3\neq 0, \text{ singular }\\ c_5\neq 0, \text{ singular } \end{aligned}$
\\
\hline
\eqref{trans-rnls-sol5}
& $\neq 0$
& $-1$
& $\neq 2$
& static
& $\begin{aligned} c_3=0, \text{ regular }\\ c_3\neq 0, \text{ singular } \end{aligned}$
\\
\hline
\eqref{scal-rnls-sol7}
& $\neq 0$
& $-1$
& $3$
& dispersive
& singular
\\
\hline
\eqref{scal-rnls-sol4}
& $\neq 0$
& $-1$
& $3$
& non-dispersive
& singular
\\
\hline
\eqref{trans-rnls-sol6}
& $\neq 0$
& $-1$
& $2$
& static
& $\begin{aligned} c_3=0, \text{ regular }\\ c_3\neq 0, \text{ singular } \end{aligned}$
\\
\hline
\end{tabular}
\end{center}
\caption{Behaviour of solutions of $n$-dimensional radial gNLS equation \eqref{nls}--\eqref{nls-dim}}
\label{classI}
\end{table}

\begin{table}[p!]
\begin{center}
\begin{tabular}{|c|c|c|c|c|c|}
\hline
solution
& $\begin{aligned} \text{nonlin.}\\\text{coeff. $k$} \end{aligned}$
& power $p\neq 0$
& source coeff. $m$
& $\begin{aligned} \text{dynamical}\\\text{behaviour} \end{aligned}$
& $\begin{aligned} \text{regularity}\\\text{at } r=0 \end{aligned}$
\\
\hline
\hline
\eqref{trans-rnls-sol7}
& $\neq 0$
& $-1$
& $-1$
& static
& regular
\\
\hline
\eqref{inver-rnls-sol4}
& $\neq 0$
& $-1$
& $-5$
& non-dispersive
& regular
\\
\hline
\eqref{inver-rnls-sol5}
& $\neq 0$
& $-1$
& $-5$
& non-dispersive
& regular
\\
\hline
\eqref{inver-rnls-sol6}
& $\neq 0$
& $-1$
& $-5$
& non-dispersive
& regular
\\
\hline
\eqref{inver-rnls-sol3}
& $< 0$
& $8/(1+\sqrt{17})$
& $(\sqrt{17}-1)/2$
& $\begin{aligned} c_2>0, \text{ dispersive }\\ c_2< 0, \text{ blow-up } \end{aligned}$
& singular
\\
\hline
\eqref{inver-rnls-sol3}
& $< 0$
& $8/(1-\sqrt{17})$
& $-(\sqrt{17}+1)/2$
& non-dispersive
& conical
\\
\hline
\eqref{scal-rnls-sol6}
& $> 0$
& $8/(1+\sqrt{17})$
& $(\sqrt{17}-1)/2$
& dispersive
& singular
\\
\hline
\eqref{scal-rnls-sol6}
& $> 0$
& $8/(1-\sqrt{17})$
& $-(\sqrt{17}+1)/2$
& dispersive
& conical
\\
\hline
\eqref{scal-rnls-sol5}
& $< 0$
& $3$
& $1/3$
& dispersive
& conical
\\
\hline
\eqref{scal-rnls-sol-r}
& $\neq 0$
& $-1$
& $-5$
& non-dispersive
& regular
\\
\hline
\eqref{scal-rnls-sol-q}
& $\neq 0$
& $-1$
& $-5$
& non-dispersive
& $\begin{aligned} c_5=0, \text{ regular }\\ c_5\neq 0, \text{ conical } \end{aligned}$
\\
\hline
\eqref{scal-rnls-sol5-apply-inver}
& $< 0$
& $3$
& $1/3$
& dispersive
& conical
\\
\hline
\eqref{scal-rnls-sol-r-apply-inver}
& $\neq 0$
& $-1$
& $-5$
& non-dispersive
& regular
\\
\hline
\eqref{scal-rnls-sol-q-apply-inver}
& $\neq 0$
& $-1$
& $-5$
& non-dispersive
& $\begin{aligned} c_5=0, \text{ regular }\\ c_5\neq 0, \text{ conical } \end{aligned}$
\\
\hline
\end{tabular}
\end{center}
\caption{Behaviour of solutions of $2$-dimensional radial gNLS equation \eqref{nls} with a point source-term}
\label{classII}
\end{table}

\begin{table}[p!]
\begin{center}
\begin{tabular}{|c|c|c|}
\hline
solution
& power $p\neq 0$
& invariance group generator
\\
\hline
\hline
\eqref{trans-rnls-sol1}
& any
& $\X_{\trans}-c_2\X_{\phas}$
\\
\hline
\eqref{trans-rnls-sol2}
& $4/n$
& ${c_2}^2\X_{\trans} + c_3c_2\X_{\scal} - k\X_{\phas} + {c_3}^2\X_{\inver}$
\\
\hline
\eqref{trans-rnls-sol2}
& $\neq 2/n,4/n$
& non-invariant
\\
\hline
\eqref{trans-rnls-sol3}
& $2/n$
& $2c_2\X_{\trans}+c_3\X_{\scal}-2k\X_{\phas}$
\\
\hline
\eqref{scal-rnls-sol2}
& $2/(2-n)$
& $2c_2\X_{\trans} +(n-4)\X_{\scal}$
\\
\hline
\eqref{trans-rnls-sol4}
& $(6-2n)/(n-2)$
& $\X_{\trans}$
\\
\hline
\eqref{scal-rnls-sol3}
& $(6-2n)/(n-2)$
& $\X_{\trans}$
\\
\hline
\eqref{trans-rnls-sol8}
& $-1$
& $\X_{\trans}+c_6\X_{\phas}$
\\
\hline
\eqref{trans-rnls-sol9}
& $-1$
& $\X_{\trans}-c_6\X_{\phas}$
\\
\hline
\eqref{trans-rnls-sol5}
& $-1$
& $\X_{\trans}$
\\
\hline
\eqref{scal-rnls-sol7}
& $-1$
& non-invariant
\\
\hline
\eqref{scal-rnls-sol4}
& $-1$
& non-invariant
\\
\hline
\eqref{trans-rnls-sol6}
& $-1$
& $\X_{\trans}$
\\
\hline
\eqref{trans-rnls-sol7}
& $-1$
& $\X_{\trans}$
\\
\hline
\eqref{inver-rnls-sol4}
& $-1$
& $\X_{\inver}$
(also $\X_{\scal}$ when $c_2=0$ or $c_3=0$)
\\
\hline
\eqref{inver-rnls-sol5}
& $-1$
& $\X_{\inver}$
\\
\hline
\eqref{inver-rnls-sol6}
& $-1$
& $\X_{\inver}$
\\
\hline
\eqref{inver-rnls-sol3}
& $8/(1\pm\sqrt{17})$
& $\X_{\inver}$
(also $\X_{\scal}$ when $c_2=0$)
\\
\hline
\eqref{scal-rnls-sol6}
& $8/(1\pm\sqrt{17})$
& $\X_{\inver}$
\\
\hline
\eqref{scal-rnls-sol5}
& $3$
& $\X_{\scal}+2c_2\X_{\inver}$
\\
\hline
\eqref{scal-rnls-sol-r}
& $-1$
& $\X_{\scal}$
\\
\hline
\eqref{scal-rnls-sol-q}
& $-1$
& $\X_{\scal}-c_2\X_{\phas}$
\\
\hline
\eqref{scal-rnls-sol5-apply-inver}
& $3$
& $\X_{\scal}+2(c_2 + c_3)\X_{\inver}$
\\
\hline
\eqref{scal-rnls-sol-r-apply-inver}
& $-1$
& $\X_{\scal}+2c_6\X_{\inver}$
\\
\hline
\eqref{scal-rnls-sol-q-apply-inver}
& $-1$
& $\X_{\scal}+2c_7\X_{\inver}-c_2\X_{\phas}$
\\
\hline
\end{tabular}
\end{center}
\caption{Symmetry invariance of radial gNLS solutions}
\label{symminv}
\end{table}

\clearpage

\section*{Acknowledgements}
S.C.A. and T.W. are each supported by an NSERC research grant. 
The work of W.F. is supported in part by the National Natural Science Foundation of China under the grant 11401529.

\end{document}